\pdfoutput=1
\documentclass[11pt,letter,subeqn,fleqn]{article}
\usepackage[normalem]{ulem}
\usepackage{amsmath, amssymb,amsthm,cite}
\usepackage[dvips]{graphics}
\usepackage[usenames,dvipsnames]{color}
\usepackage{graphicx}
\usepackage{tabulary}
\usepackage{float}
\usepackage{gensymb}
\usepackage{authblk}
\usepackage{caption}
\usepackage{subcaption}

\title{Radial waves in fiber-reinforced axially symmetric hyperelastic media}

\author[1]{Alexei Cheviakov\thanks{Corresponding Author. Alternative English spelling: Alexey Shevyakov. Electronic mail: alexei.cheviakov@usask.ca}}
\author[1]{Caylin Lee}
\author[2]{Rehana Naz \thanks{Electronic mail: drrehana@lahoreschool.edu.pk}}
\affil[1]{Department of Mathematics and Statistics, University of Saskatchewan}
\affil[2]{Centre for Mathematics and Statistical Sciences, Lahore School of Economics}

\def\beq{\begin{equation}}
\def\eeq{\end{equation}}
\def\barr{\begin{array}{ll}}
\def\earr{\end{array}}

\def\sg#1{{\rm #1}}
\def\const{\hbox{\rm const}}

\def\Tr{\mathop{\hbox{\rm Tr}}}
\def\grad{{\hbox{\rm grad}}}
\def\div{{\hbox{\rm div}}}

\def\tens#1{{\boldsymbol{\rm #1}}} 

\def\vec#1{{\boldsymbol{\rm #1}}} 
\def\tens#1{{\boldsymbol{\rm #1}}} 

\newcommand\pder[2][]{\ensuremath{\dfrac{\partial #1}{\partial #2}}}

\newtheorem{theorem}{Theorem}

{\theoremstyle{definition}

}

\newcounter{tabnum}

\renewcommand{\div}{\mathrm{div}}

\newcommand{\C}{\mathbf{C}}

\begin{document}

\maketitle \numberwithin{equation}{section}
\maketitle \numberwithin{remark}{section}
\numberwithin{lemma}{section}
\numberwithin{proposition}{section}

\begin{abstract}

Complex elastic media such as biological membranes, in particular, blood vessels, may be described as fiber-reinforced solids in the framework of nonlinear hyperelasticity. Finite axially symmetric anti-plane shear displacements in such solids are considered. A general nonlinear wave equation governing such motions is derived. It is shown that in the case of Mooney-Rivlin materials with standard quadratic fiber energy term, the displacements are governed by a linear cylindrical wave equation.

Extensions of the model onto the case when fibers have a radial projection, as well as onto a viscoelastic case taking into account dissipative effects, are considered; wave equations governing shear displacements in those cases are derived and analyzed.


\end{abstract}

\section{Introduction}

The framework of nonlinear elasticity is commonly used to model finite deformations of elastic materials, with applications ranging from industry and construction to biological tissues, membranes, and cell biology. Mathematical foundations of elastodynamics theory date back to Hooke, Navier and Cauchy; in its current form, in particular, in the language of geometry, much of the theory has been formulated in the second half of the 20th century (see, e.g., Refs.~\cite{Truesdell1965, Truesdell1968, Wang1973, ciarlet1988mathematical, marsden1994mathematical, holzapfel2000nonlinear, Chadwick2012} and references therein). Multiple extensions of elasticity theory allow to take into account anisotropic, viscous, elastoplastic, thermal, and other physical effects \cite{Ivlev66, truesdell_rational_1969, Ziegler83, khan1995continuum,Rajagopal98, lubarda2002elastoplasticity, hutter_continuum_2004, Wu05, Wang2012,cheviakov2013finite}. The problem of choice of an appropriate framework and constitutive relationships, as well as the algebraic complexity and the essentially nonlinear nature of the governing equations themselves, remain among the main challenges in solid mechanics-based modeling.

In contrast with approximate models based on linear approximations (incremental analysis), a continuum model that strives to take into account all important physical laws governing a process of interest is usually given by a system of nonlinear partial differential equations (PDE). The set of dependent variables commonly includes the Eulerian coordinates, or, equivalently, finite (non-small) displacements of material points, and other physical fields (e.g., \cite{Fu2001, Ogden2007}). While a significant body of theoretical results is available in the field of nonlinear elastostatics, for time-dependent problems, nonlinear effects such as instability, non-existence or non-uniqueness of solutions, existence of multiple scales, shocks, finite-time blowup, etc., lead to significant complications in obtaining exact or approximate solutions and analysis of solution behaviour. The study of wave propagation in nonlinear elastic media is an active research area \cite{Destrade2005, Ogden2007,Saccomandi2007,Braun2007, Marasco2009, merodio2007rectilinear, cheviakov2015fully, cheviakov2016one}, with multiple applications to the study of biological materials (e.g., \cite{pioletti2000non, merodio2007thermodynamically, ogden2006mechanics, destrade2007creep, Kalita2008, Namani2009, destrade2009inhomogeneous, Umale2013, Saez2014}), medical imaging (e.g., \cite{parker1992sonoelasticity, Sandrin2003, Valdez2013}), geosciences, and other areas.

The current paper is concerned with the investigation of certain reductions of fully nonlinear models of anisotropic fiber-reinforced elastic and viscoelastic solids, leading to nonlinear wave equations. The framework of hyperelasticity and viscoelasticity and its generalizations has recently been actively used to model multiple types of media,  including fabrics and biological materials \cite{peng2010simple, gasser2006hyperelastic, ehret2007polyconvex, marchesseau2010fast, roan2011strain, pena2011formulation, cheviakov2015fully}. Equations of nonlinear hyperelasticity are based on the existence of a potential (the stored energy function) which defines local stresses based on local displacements of material particles. The resulting equations for Eulerian coordinates of material particles essentially have the form of Newton's second law, expressing the local balance of forces, and can be viewed as a system of coupled wave equations in 3+1 dimensions. The form of the stored energy function, initial density and stress distribution, and external forces determine the mechanical properties of a given configuration. The choice of constitutive functions can be based on various considerations; constitutive modeling of complex media is an active research area by itself (see, e.g., \cite{Gent1996, GangMR2010, assidi2011equivalent, Cheviakov2012, cheviakov2018symbolic, hess2018solution} and references therein). The presence of elastic fibers, for example, in biological tissues and many man-made materials, alters the mechanical response of the elastic substance. In order to incorporate fiber stretch and interaction effects into the hyperelasticity framework, fiber-dependent terms are added to the stored energy function; multiple constitutive models have been developed for such terms. Some models are discussed, for example, in Refs.~\cite{holzapfel2000new, Horgan2005b, cheviakov2015fully, cheviakov2016one}.

In many cases, mechanical behaviour of an elastic material can be modeled with high accuracy as {incompressible}, or volume-reserving. This assumption simplifies the transformation between Lagrangian and Eulerian coordinates, which in this case has a unit Jacobian, and often makes the final PDEs significantly simpler \cite{le1993three, Horgan2005b, Destrade2013, merodio2005mechanical, Cheviakov2012}.

Due to the complexity of full three-dimensional equations and boundary problems required to model, for example, the elastic behavior of a human organ, it is usually not feasible to derive exact or approximate closed-form solutions or obtain other useful properties of such models; numerical simulations remain the common avenue. To gain insights into the mechanical processes, reductions based on symmetries and/or other assumptions are a common way to proceed \cite{BCABook}. It is known that specific settings and reduction ans\"{a}tze in nonlinear elastodynamics can yield scalar coupler or decoupled equations describing the propagation of certain perturbations (e.g., \cite{destrade2010scalar, cheviakov2015fully}). In the framework of incremental analysis, such models are linear, whereas the finite elasticity theory leads to fully nonlinear models. Such single hyperbolic wave-type equations are of interest from both physical and mathematical points of view. For linear wave equations, with constant or variable coefficients, a large set of classical tools is available, such as Fourier and Green's function methods, or the method of characteristics \cite{Zauderer}; for nonlinear models, the situation is significantly more complex. Some of such models are integrable, being, for example, exactly linearizable through hodograph-type or nonlocal transformations \cite{BCABook}. However, this is not the case for the majority of nonlinear wave equations, where nonlinear effects can lead to loss of regularity or hyperbolicity, shock formation, etc.~\cite{merodio2007rectilinear, destrade2009inhomogeneous, cheviakov2015fully}. Dissipative, in particular, viscoelastic effects, which may regularize the model, should be taken into account in such cases \cite{Kwon1996, Magnenet2007, cheviakov2016one}. Non-hyperbolic, for example, evolution equations, and more complex PDEs describing the propagation of perturbations, have been shown to arise in various mechanical contexts, in particular, in the study of shear waves in incompressible solids \cite{destrade2010scalar}.

In this work, we apply the framework of full three-dimensional anisotropic incompressible hyperelasticity and visco-hyperelasticity to study shear waves in cylindrical geometry. The study is motivated by both industrial and biomedical applications \cite{cheviakov2016one}. It is well known that walls of blood vessels, in particular, arteries, are multi-layered structures, with each layer having its own mechanical properties \cite{holzapfel2000new, Basciano2009}. From the mechanical point of view, the two most significant arterial layers  are adventitia and media, each containing two sets of helically oriented collagen fibers \cite{holzapfel2000new, gasser2006hyperelastic}. In Ref.~\cite{cheviakov2016one}, nonlinear wave equations corresponding to finite shear displacements in a medium with one and two embedded fiber families were derived; in particular, a ``flat cylinder" model approximating an arterial wall was used, based on Cartesian coordinates. The current work extends the results of Ref.~\cite{cheviakov2016one}, taking into account the cylindrical geometry from the very beginning.

The paper is organized as follows. The general physical model setup in cylindrical geometry and the necessary components of the framework of incompressible finite hyperelasticity, constitutive modeling, fiber-related anisotropy, and viscoelastic effects are introduced in Section \ref{Sec:IncompressibleFiniteElasticity}. In Section \ref{Sec:HyperelasticLinearModel}, the propagation of s-waves in the radial direction is considered for cylindrical media with two embedded helically directed sets of identical fibers making the same pitch angle but opposite chiralities. This ansatz is inherently incompressible. We show that when the stored energy function is a sum of two arbitrary smooth components, responsible respectively for isotropic and fiber-related effects, the vertical displacements $G(t,R)$ of material points located at material radii $R$ are governed by a single PDE in the divergence form:
\beq\label{G:for:genW:f}
G_{tt} = \dfrac{1}{R}\dfrac{\partial }{\partial R}\left( R\, f(G_R)\right),
\eeq
where $f$ is a function of $G_{R}$ that depends on the form of the stored energy function, and the second unknown, the hydrostatic pressure $p(t,R)$, is expressed in terms of $G(t,R)$. [In \eqref{G:for:genW:f} and below, where appropriate, partial derivatives are denoted by subscripts: $G_{R}={\partial G}/{\partial R}$, $G_{tt}={\partial^2 G}/{\partial t^2}$, etc.] Moreover, in the important case when the stored energy function is a combination of the incompressible Mooney-Rivlin isotropic part and a standard quadratic reinforcement anisotropic term, as well as in several other cases, the displacements $G(t,R)$ of material points are shown to satisfy a linear PDE
\beq\label{G:lin}
G_{tt} = \alpha \left(G_{RR} + \dfrac{1}{R}G_R\right),
\eeq
where $\alpha = \const$ is a fiber-independent material parameter. The wave equation \eqref{G:lin} is a well-known equation describing, for example, axially symmetric modes of small oscillations of an elastic circular membrane \cite{Zauderer}. It is therefore shown that in such cases, radial shear waves do not ``feel" the presence of fibers which are tangent to cylinders $R=\const$. Physical boundary value problems for the linear and the nonlinear wave equations are discussed; in particular, a boundary value problem for the PDE \eqref{G:lin} with common boundary conditions corresponding to stationary, free, or forced boundaries of the cylindrical domain $R_1 \leq R\leq R_2$ can be solved explicitly by separation of variables.

It turns out that when the fibers in the medium are not \emph{exactly} helical, specifically, when they have a nonzero projection on the radial direction, then in the same hyperelasticity framework, for the same constitutive model that leads to \eqref{G:lin}, shear waves are described by nonlinear equations. In Section \ref{sec:4:modif:fibers}, we consider such a modified-fiber model, with radial fiber projections measured by an angle parameter $\delta\ne 0$. We show that in this case, the displacements are governed by a family of nonlinear wave equations
\beq\label{G:4N}
G_{tt} = \dfrac{1}{R} \dfrac{\partial}{\partial R}  \left[ R\left( N_1G_{R}+  N_2 G_R^2 + N_3 G_R^3 + N_4\right)\right]\,,
\eeq
where $N_1, \ldots, N_4$ are constant parameters depending on fiber angles, and mechanical properties. The PDEs \eqref{G:4N} belong to the class \eqref{G:for:genW:f}. In particular, for small $\delta$, the coefficients $N_1=\alpha + \mathcal{O}(\delta^2)$, and $N_i=\mathcal{O}(\delta^{n_i})$, $n_i\geq 1$, $i=2,3,4$, consistent with the PDE \eqref{G:lin} in the limit $\delta\to 0$. The nonlinear wave equations \eqref{G:4N} with polynomial nonlinearities have not, to our knowledge, been studied in detail in the literature; in particular, they are not known to be linearizable by a local or nonlocal transformation. Equivalence transformations and point symmetries of an even more general family of wave equations
\[
u_{tt}=f(x,u_x) u_{xx}+g(x,u_x)
\]
for the unknown $u=u(x,t)$, with two arbitrary functions $f,g$, have been systematically classified in Ref.~\cite{bihlo2012complete} (see also references therein), but cases relevant to elasticity problems that arise in the current study have not been specifically considered.

Non-dissipative mechanical systems commonly admit a classical Lagrangian; it is shown that the symmetry-reduced wave equations \eqref{G:for:genW:f} also arise from a variational principle (Section \ref{sec:4:modif:fibers}). Sample numerical solutions of  the PDE \eqref{G:4N} corresponding to unidirectional waves are presented and shown to develop ``corners" as a consequence of the nonlinearity (cf. \cite{cheviakov2016one}).

Finally, in Section \ref{sec:visc}, the anti-plane shear helical fiber wave model of Section \ref{Sec:HyperelasticLinearModel} is amended with viscoelastic effects incorporated into the strain energy density through pseudo-invariants that involve time derivatives of the Cauchy-Green stress. It is shown that instead of the linear wave equation \eqref{G:lin}, for this model, the displacement $G(t,R)$ of the shear wave is described by a third-order scalar PDE
\beq \label{Geq:visco}
G_{tt} = \dfrac{1}{R}\dfrac{\partial}{\partial R}\Big(R G_R\left[\alpha+\mu_1 G_R G_{tR}\left(1+2 G_R^2\right)\right]\Big)\, ,
\eeq
with a viscosity-related coefficient $\mu_1$. Numerical simulations show that the viscosity term provides a regularization-type effect.

The paper is concluded with a discussion in Section \ref{sec:Discussion}.



\section{The Incompressible Finite Hyperelasticity Framework}\label{Sec:IncompressibleFiniteElasticity}

The fully nonlinear hyperelasticity framework considers finite (as opposed to infinitesimally small) displacements of solid elastic bodies. We briefly review the notation and the main elements of mathematical models in incompressible hyperelasticity. Boldface notation is used  for vector and tensor quantities. Partial derivatives will often be denoted by subscripts: $\partial f /{ \partial t}\equiv f_t$, etc. We also assume summation in repeated indices where appropriate.

\begin{figure}[htbp]
  	\begin{center}
		\includegraphics[width = 0.8\textwidth]{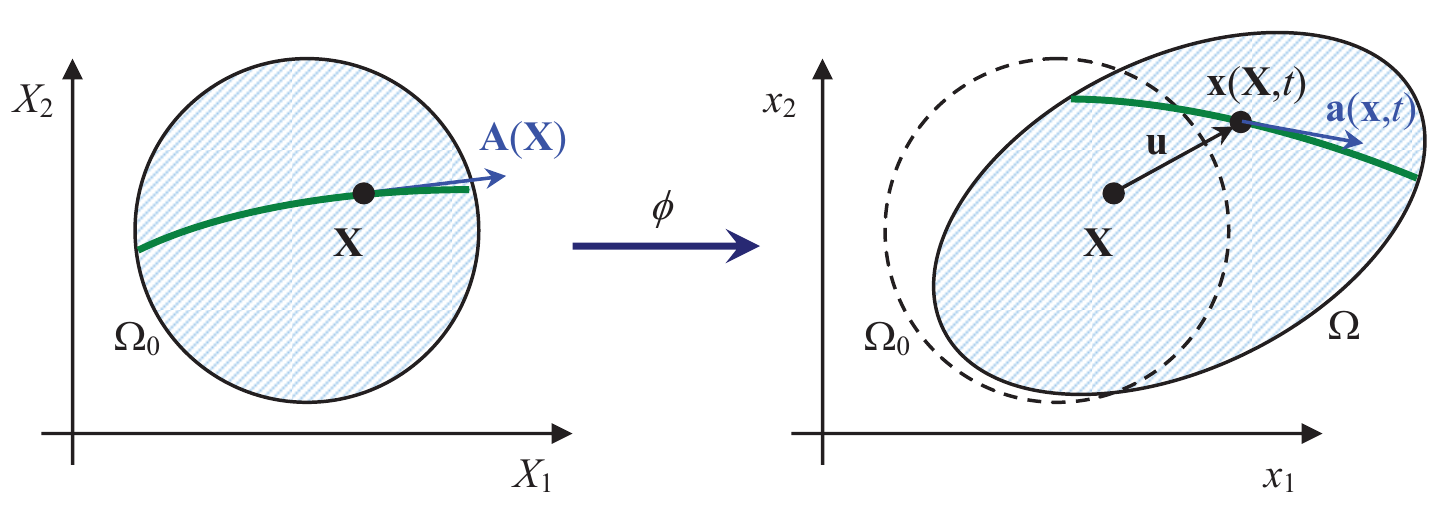}
  	\end{center}
    \caption{ \label{fig:main} The material (Lagrangian) and the actual (Eulerian) domains, Lagrangian and Eulerian coordinates, the displacement, and the fiber direction vectors (see Section \ref{sec:fiberintro}).}
\end{figure}

\medskip
Consider a solid body that at the current time $t$ occupies a spatial domain $\overline{\Omega}\subset \mathbb{R}^3$. The actual positions of material points in the body (the Eulerian coordinates) are given by
\beq\label{eq:Xtox}
\vec{x} = \vec{\phi}(\vec{X}, t) = \vec{X} + \vec{u},
\eeq
where $\vec{X}$ are material coordinates (or Lagrangian coordinates, the labels of the material points), and $\vec{u}=\vec{u}\left(\vec{X}, t\right)$ denotes the displacement of a material point labelled by $\vec{X}$. The material coordinates are often taken to be initial conditions: $\vec{\phi}\left(\vec{X}, 0\right)=\vec{X}$. The material coordinates run through the spatial region ${\Omega}_0\subset \mathbb{R}^3$, called the reference, the material, or the Lagrangian configuration, whereas the actual domain is given by ${\Omega}= \vec{\phi}({\Omega}_0)$ (Figure \ref{fig:main}). In the fully nonlinear framework, since the displacements $\vec{u}$ are not assumed to be small, the equations of motion are commonly written in terms of actual particle positions $\vec{x}$.

The velocity of a material point ${\bf X}$ is given by
\[
\vec{v}(\vec{X},t)=\frac{d\vec{x}}{dt}=\frac{d\vec{u}}{dt}.
\]
The mapping \eqref{eq:Xtox} is assumed to be invertible, sufficiently smooth, and physical. In particular, the deformation gradient provided by the Jacobian matrix
\beq\label{eq:Fdef}
\tens{F}(\vec{X},t) = \grad_{(\vec{X})} \,\vec{\phi},\qquad F^i_{~j} = \frac{\partial x^i}{\partial X^j}=F_{ij}
\eeq
is invertible, and satisfies the orientation-preserving condition
\[
J =\det {\tens{F}} > 0.
\]
(For Cartesian coordinates and the flat space metric $g^{ij}=\delta^{ij}$, the indices of all tensors can be freely raised or lowered.) The related symmetric left and right Cauchy-Green strain tensors $\tens{B}$ and $\mathbf{C}$ given by
\beq\label{eq:BC}
\tens{B} = \tens{F}\tens{F}^T,\qquad \tens{C} = \tens{F}^T\tens{F}
\eeq
play an important role in solid mechanics. If the density of the elastic substance in the reference configuration is denoted by $\rho_0=\rho_0(\vec{X})$, the actual time-dependent density in Eulerian coordinates takes the form
\[
\rho(\vec{X},t)=\rho_0/J.
\]
For incompressible materials, one has
\[
J=1,\qquad \rho(\vec{X},t)=\rho_0(\vec{X}).
\]
For the  applications considered in this work, we will take the material density $\rho_0=\const$, however, general formulas within Section \ref{Sec:IncompressibleFiniteElasticity} hold for an arbitrary $\rho_0(\vec{X})$.

\subsection{Equations of motion of a hyperelastic material}

According to the Cauchy theorem, in the Eulerian configuration, the force acting on a unit surface area with a unit normal $\vec{n}$ within the elastic solid,  $\vec{t}=\tens{\sigma}\vec{n}$, is expressed in terms of the symmetric Cauchy stress tensor $\tens{\sigma}$. Similarly, the force acting on a surface element with the unit normal $\vec{N}$ in the Lagrangian configuration is given by $\vec{T}=\tens{P}\vec{N}$, where $\tens{P}$ is the non-symmetric first Piola-Kirchhoff tensor, related to the Cauchy stress tensor through
\beq\label{eq:stress:PK_Cauchy}
\tens{P}=J\tens{\sigma}\tens{F}^{-T},
\eeq
here $\tens{F}^{-T}$ denotes the transpose of the inverse of the deformation gradient. The related second Piola-Kirchhoff tensor is given by $\tens{S}=\tens{F}^{-1}\tens{P}$.

For hyperelastic materials, the forms of Piola-Kirchhoff stress tensors $\tens{P}$, $\tens{S}$ follow from a postulated form of a ``stored energy" function $W^h$ \cite{Marsd}, which is the strain energy per unit mass. [The volumetric strain energy density in the material frame of reference is given by $\rho_0 W^h$.] For isotropic hyperelastic media, $W^h= W^h\left({\bf X},{\bf F}\right)$. For anisotropic  materials involving fibers, $W^h = W^h\left(\vec{X},{\bf F}, \vec{A}_1,\ldots,\vec{A}_k\right)$, where the unit vectors ${\vec{A}_j}$, $j=1,\ldots,k$ define the direction fields of $k$ independent, possibly interacting fiber families in the reference configuration. For an anisotropic materials, it is common to state the strain energy density as a sum of an isotropic and an anisotropic contribution (e.g., \cite{Basciano2009}):
\beq\label{new_W_gen}
W^h = W^h_{\rm iso} + W^h_{\rm aniso}.
\eeq
Such a potential energy function is assumed to fully describe the material behavior. The choice of a specific form of $W^h$ for the given material and physical situation is the main problem of \emph{constitutive modeling} (e.g., \cite{cheviakov2018symbolic, cheviakov2015fully,cheviakov2016one, ciarlet1988mathematical} and references therein).

For incompressible models, the form of the stored energy  $W^h$ yields the Piola-Kirchhoff stress tensors through the formulas
\beq\label{eq:PK1:from:Wh}
\tens{P} = -p\;\tens{F}^{-T} + \rho_0\dfrac{\partial W^h}{\partial\tens{F}} = \tens{F}\,\tens{S},\qquad
\eeq
\beq\label{eq:PK2:from:Wh}
\tens{S} = -p\;\tens{C}^{-1} + 2 \rho_0 \,\dfrac{\partial W^h}{\partial\tens{C}},
\eeq
measured in the units of pressure; here $p=p(\vec{X},t)$ is the hydrostatic pressure. Since $\tens{C}$ is symmetric, the formula \eqref{eq:PK2:from:Wh}
is understood in the sense
\beq\label{c:symm}
\dfrac{\partial W^h}{\partial\tens{C}}\equiv \dfrac{1}{2}\left(\dfrac{\partial W^h}{\partial\tens{C}}+\dfrac{\partial W^h}{\partial\tens{C}^T}\right).
\eeq
Consequently, if the strain energy density is expressed purely as a function of the components of $\tens{C}$, then the expression $W^h=W^h(\tens{C})$ is symmetrized by the substitution
\[
C_{ij}\to \dfrac{1}{2}(C_{ij}+C_{ji}).
\]


The dynamics of an incompressible hyperelastic solid is described by an initial-boundary value problem for a set of equations of motion. The latter are given by the momentum conservation and the incompressibility condition:
\begin{subequations} \label{eq:motion:basic:lagr}
\beq\label{eq:motion:basic:lagr:mom}
\rho_0\vec{x}_{tt} = \div_{(\vec{X})}\tens{P}+\vec{Q}, 
\eeq
\beq\label{eq:motion:J1}
1-J =0. 
\eeq
\end{subequations}
The vector PDE \eqref{eq:motion:basic:lagr:mom} is a variant of Newton's second law, expressing the conservation of momentum in the reference configuration. In \eqref{eq:motion:basic:lagr}, $\vec{Q}=\vec{Q}(\vec{X},t)$ is the total external body force per unit volume, and the divergence of $\tens{P}$ with respect to the material coordinates is given by
\[
(\div_{(\vec{X})}\tens{P})^i=\frac{\partial P^{ij}}{\partial X^j}.
\]
In addition, the motions are required to satisfy the condition
\beq\label{eq:motion:basic:lagr:symm}
\tens{F}\tens{P}^T=\tens{P}\tens{F}^T.
\eeq
which expresses the conservation of angular momentum, and is equivalent to the Cauchy stress tensor symmetry requirement $\tens{\sigma}=\tens{\sigma}^T$. For isotropic materials, as well as in some other cases, this symmetry condition is identically satisfied (e.g., \cite{Cheviakov2012}).






The equations of motion \eqref{eq:motion:basic:lagr} are formulated in the Lagrangian framework, with independent variables $(\vec{X},t)$, but they can also be written in the laboratory (Eulerian) frame of reference, with independent variables $(\vec{x},t)$ and dependent variables $\vec{v}(\vec{x},t)$. The Eulerian form of the governing equations is commonly used in fluid dynamics.

The incompressible model \eqref{eq:motion:basic:lagr} is relevant in the physical space $\mathbb{R}^n$, $n=2,3$, as well as in compatible symmetry-reduced settings, but in the Cartesian one-dimensional case $n=1$, the incompressibility requirement $J=1$ is overly restrictive, allowing only for translation-type motions $x=X+a(t)$.

When the external forces vanish or are potential forces, the general three-dimensional equations of motion \eqref{eq:motion:basic:lagr} of a hyperelastic solid admit a variational formulation. In particular, the PDEs are obtained from the variation of the action functional
\beq\label{rem:SLagr}
\mathcal{S}=\int_{0}^{\infty} \int_{\mathbb{R}^n} \mathcal{L} \, d^n x\, dt,
\eeq
where the Lagrangian density is given by
\beq\label{rem:LLagr}
\mathcal{L} = \rho_0 (W-K) + p(1-J).
\eeq
For the case of no external forces $\vec{Q}=0$, the potential energy is the hyperelastic strain energy per unit mass $W= W^h\left({\bf X},{\bf F}\right)$ \eqref{new_W_gen}, and
\[
K=\dfrac{1}{2} \sum_{i=1}^{n} (x^i_t)^2
\]
is the kinetic energy per unit mass. The Euler operator with respect to $u$ is  defined as
\[
\sg{E}_{u} =
\frac{\partial }{\partial u} - \sg{D}_i \frac{\partial}{\partial u_i}
+ \cdots
+ (-1)^l\sg{D}_{i_1 }\ldots\sg{D}_{i_l} \frac{\partial}{\partial u_{i_1 \ldots i_l}} + \cdots,
\]
where $u$ is any scalar dependent variable, $u_i$ denotes its derivative by $i$-th independent variable, and $\sg{D}_i$ is the corresponding total derivative operator. Then the extremals of \eqref{rem:SLagr} satisfy the Euler-Lagrange equations $\sg{E}_{u} \mathcal{L}=0$ for $u=p$, $x^1$, $x^2$, $x^3$. These equations are indeed the PDEs \eqref{eq:motion:basic:lagr} as they stand:
\[
\begin{array}{lllll}
\dfrac{\delta \mathcal{L}}{\delta p} &\equiv & \sg{E}_p \mathcal{L} &=& 1-J =0,\\[2ex]
\dfrac{\delta \mathcal{L}}{\delta x^k} &\equiv& \sg{E}_{x^1} \mathcal{L} &=&\rho_0x^k_{tt} - \dfrac{\partial P^{kj}}{\partial X^j} =0,\quad k=1,2,3.
\end{array}
\]
The variational formulation of a nonlinear model is a useful property; in particular, it yields a direct relation between local variational symmetries and conservation laws of a model through the first Noether's theorem (e.g., \cite{Olver}), and may be related with integrability.  Based on the existence of a variational formulation for a general model, there, however, is no straightforward statement about the existence of a variational formulation for a reduced model; moreover, the variational property is rather ``unstable" with respect to various transformations \cite{BCABook,BCApap}. It is, however, possible to show that the reduced wave model and its extensions considered in this paper (Sections \ref{Sec:HyperelasticLinearModel} and \ref{sec:4:modif:fibers} below) also admit a variational formulation.

\subsection{Constitutive models in isotropic and anisotropic hyperelasticity}\label{sec:fiberintro}

A constitutive relation for an isotropic homogeneous hyperelastic material is commonly posed as an expression of the strain energy density $W^h=U(I_1, I_2, I_3)$ in terms of the principal invariants of the  Cauchy-Green strain tensors $\tens{B}$ and $\mathbf{C}$ \eqref{eq:BC}:
\beq\label{eq:invB}
I_1=\Tr\tens{C}, \qquad I_2=\frac{1}{2}[(\Tr \tens{C})^2-\Tr
(\tens{C}^2)],\qquad  I_3=\det \tens{C} = J^2.
\eeq
For incompressible materials, $J=1$, hence generally, one has $W^h_{\rm iso} =U(I_1,I_2)$. Since in the natural state $\vec{x}=\vec{X}$, both invariants $I_1=I_2=3$, and the general isotropic constitutive relation is commonly written as
\beq\label{eq:constit:iso}
W^h_{\rm iso} =U(I_1,I_2).
\eeq
In addition to any differentiability requirements on the function $U$, for non-prestressed configurations, it also must satisfy the physical condition of the natural state: if $\vec{x}=\vec{X}$, in other words, all displacements are zero, then
\beq\label{eq:constit:natural} 
\dfrac{\partial U}{\partial I_1} +2\dfrac{\partial U}{\partial I_2}=0
\eeq
(see, e.g., \cite{Cheviakov2012}.) If pre-stressed configurations are allowed, the condition \eqref{eq:constit:natural} may not hold.

Multiple constitutive models \eqref{eq:constit:iso} have been suggested for specific applications, involving, for example, polynomial and exponential-type forms of the isotropic stored energy function $U$; for a review, see, e.g., \cite{thesisBader, merodio2007thermodynamically, Cheviakov2012, cheviakov2015fully}. A wide class of rubber-like materials is described by the Mooney-Rivlin constitutive relation
\beq\label{eq:constit:W_MR}
U(I_1,I_2) = a(I_1-3)+b(I_2-3),
\eeq
with material parameters $a,b=\const>0$. It corresponds to the lowest-order terms of a series expansion of a general analytic function $U(I_1,I_2)$. A simpler case $b=0$ is the neo-Hookean model.

\medskip Anisotropic hyperelastic materials with fibers are modeled using a stored energy contribution $W^h_{\rm aniso}$, commonly assumed to depend on pseudo-invariants involving fiber directions, and the corresponding fiber strength and interaction parameters. Each fiber family is given by a vector field $\vec{A}_j=\vec{A}_j(\vec{X})$, $|\vec{A}_j|=1$, at every point of the material configuration. [For the purposes of formula presentation, the direction fields $\{{\vec{A}_j}\}_{j=1}^k$ are assumed to be  column vectors.] In the Eulerian frame of reference, the time-dependent fiber orientation fields for each family of fibers are determined by
\beq\label{eq:stretch}
\lambda_j \vec{a}_j=\vec{F}\vec{A}_j,\quad j=1,\ldots,k,
\eeq
where $\vec{a}_j=\vec{F}\vec{A}_j/|\vec{F}\vec{A}_j|$ are unit fiber direction vectors in the Eulerian configuration (see Figure \ref{fig:main}), and $\lambda_j=|\vec{F}\vec{A}_j|$ are the stretch factors.

For a single fiber family, $k=1$, the two pseudo-invariants are given by
\beq\label{eq:aniso invar1}
I_4 = \vec{A}_1^{T} \C \vec{A}_1, \qquad I_5 = \vec{A}_1^{T} \C^{2} \vec{A}_1.
\eeq
where $I_4\equiv \lambda_1^2$ is the squared fiber stretch factor, and $I_5$ relates to the effect of the fiber on the shear response in the material \cite{Demi2010,merodio2005mechanical,pandolfi2006model, cheviakov2015fully}. (In particular, if $\vec{w}=\tens{C} \vec{A}$ is a push-forward of the material vector $\vec{A}$ by the right Cauchy-Green tensor then $I_5 = \vec{w}^T\vec{w} = |\vec{w}|^2$.) A stored energy function that takes into account these effects takes the general form
\[
W^h_{\rm aniso}=V(I_4,I_5).
\]

For the case of a material with two fiber families given by direction vectors $\vec{A}_1$ and $\vec{A}_2$, the anisotropic stored energy part generally depends on five pseudo-invariants \cite{holzapfel2000nonlinear}:
\beq\label{eq:Eaniso:2fib:gen}
W^h_{\rm aniso}=V(I_4, I_5, I_6, I_7, I_8).
\eeq
Here $I_6$ and $I_7$ given by
\beq\label{eq:aniso invar2}
I_6 = \vec{A}_2^{T} \C \vec{A}_2, \qquad I_7 = \vec{A}_2^{T} \C^{2} \vec{A}_2,
\eeq
have are the same meaning as \eqref{eq:aniso invar1} for the second fiber family, and an additional pseudo-invariant
\beq\label{eq:aniso invar:int}
I_8 = ( \vec{A}_1^T \vec{A}_2)\vec{A}_1^{T} \C \vec{A}_2
\eeq
describes coupling between the fiber families. It has been used, for example, in modeling of cornea \cite{pandolfi2006model}.



\subsection{Hyper-viscoelastic constitutive models}\label{sec:visco:framew}

In many materials, viscoelastic, as opposed to hyperelastic, behaviour is exhibited. Various approaches exist for the mathematical description of viscoelasticity, including rational and irreversible thermodynamics, finite viscoelasticity, and hyper-viscoelasticity. For a more detailed review, see, e.g., Refs.~\cite{thesisBader, cheviakov2016one}, and references therein.

In the current contribution, we use the hyper-viscoelasticity framework \cite{holzapfel2000nonlinear}, which employs a hyperelastic stored energy part $W^h$ \eqref{new_W_gen} to describe elastic effects, and a  ``dissipative potential" $W^v$ associated to the viscous phenomena. The second Piola-Kirchhoff tensor formula \eqref{eq:PK2:from:Wh} is modified to include the viscoelastic stress
\beq\label{eq:PK2:viscoel}
\tens{S}_v = 2 \rho_0 \,\dfrac{\partial W^v}{\partial\tens{\dot{C}}}\,.
\eeq
Using \eqref{eq:PK2:from:Wh}, one has the total stress tensor expression
\beq\label{eq:PK2:viscoel:tot}
\tens{S} = \tens{S}_h + \tens{S}_v = -p\;\tens{C}^{-1} + 2 \rho_0 \left(\dfrac{\partial W^h}{\partial\tens{C}} + \dfrac{\partial W^v}{\partial\tens{\dot{C}}}\right),
\eeq
where \eqref{c:symm} is taken into account. The equations of motion of the solid are still given by \eqref{eq:motion:basic:lagr}, with $\tens{P} = \tens{F}\,\tens{S}$.

The forms of both the hyperelastic stored energy $W^h$ and the dissipative potential $W^v$ vary by the application; for  example, the viscoelastic model of a fiber-reinforced material with a single fiber family studied by Pioletti and Rakotomanana \cite{pioletti2000non} uses the following energy density expressions:
\beq\label{eq:MerG:Wh}
W^h=\dfrac{\mu}{2} (I_1-3) + \dfrac{k_1}{2k_2}\left(e^{k_2(I_4-1)^2}-1\right),
\eeq
\beq\label{eq:MerG:Wv}
W^v=\dfrac{\eta_1}{4} J_2 (I_1-3) + \eta_2 J_9\dfrac{k_1}{2k_2}\left(e^{k_2(I_4-1)^2}-1\right),\qquad \eta_1, \eta_2=\const,
\eeq
where  $\mu, k_1, \eta, \gamma$ are the appropriate dimensional constant parameters, $k_2$ is a dimensionless constant, and
\beq\label{eq:J2J9}
J_2 = \rm{Tr}(\tens{\dot{C}^2}),\qquad J_{9} = \vec{A}^T \tens{\dot{C}^2} \vec{A}
\eeq
are the corresponding viscoelastic pseudo-invariants. Other pseudo-invariants $J_1$, $\ldots$, $J_{13}$ have been constructed and used in the case of a single fiber family \cite{merodio2007thermodynamically}; their set is naturally extended for multiple fiber bundles.

In Ref.~\cite{cheviakov2016one}, for two fiber families given by material directions $\vec{A}_1$, $\vec{A}_2$, the viscoelastic strain energy form
\begin{equation}\label{eq:W:visc:ours}
W^v = \frac{\mu_1}{4} J_2 \left(I_1-3\right) + \frac{\mu_2}{2} J_{9,1} \left(I_4-1\right)^2 + \frac{\mu_3}{2}J_{9,2}\left(I_6 -1\right)^2\, ,
\end{equation}
was used, with material viscosity parameters $\mu_i$, $i=1,2,3$. Here $J_{9,1}$ and $J_{9,2}$ denote the pseudo-invariant $J_{9}$ \eqref{eq:J2J9} computed respectively for $\vec{A}_1$ and $\vec{A}_2$. The viscoelastic potential \eqref{eq:W:visc:ours} corresponds to the leading Taylor terms of \eqref{eq:MerG:Wv}, adapted to include two fiber families.

A general class of viscoelastic strain energy expressions for two fiber families, depending on the same above pseudo-invariants, is given by
\begin{equation}\label{eq:W:visc:ours:genfam}
W^v = W(I_1, I_2, I_4, I_5, I_6, I_7, I_8, J_{9,1}, J_{9,2}).
\end{equation}

%


\section{Radial shear waves in a cylindrical hyperelastic solid with two helical fiber families}\label{Sec:HyperelasticLinearModel}

\begin{figure}[htbp]
\centering
        \begin{subfigure}[b]{0.3\textwidth}
                \centering
                \includegraphics[width=\textwidth]{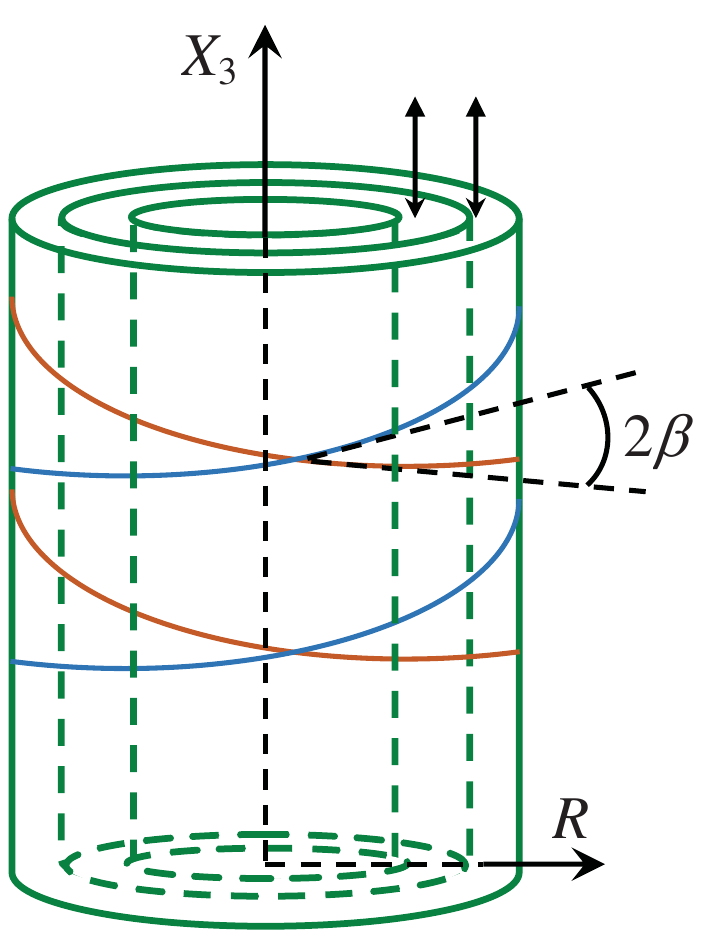}
                \caption{}
                \label{fig:gull}
        \end{subfigure}%

\begin{minipage}{.48\textwidth}
  \centering
        \begin{subfigure}[b]{\textwidth}
                \centering
                  \includegraphics[width=\textwidth]{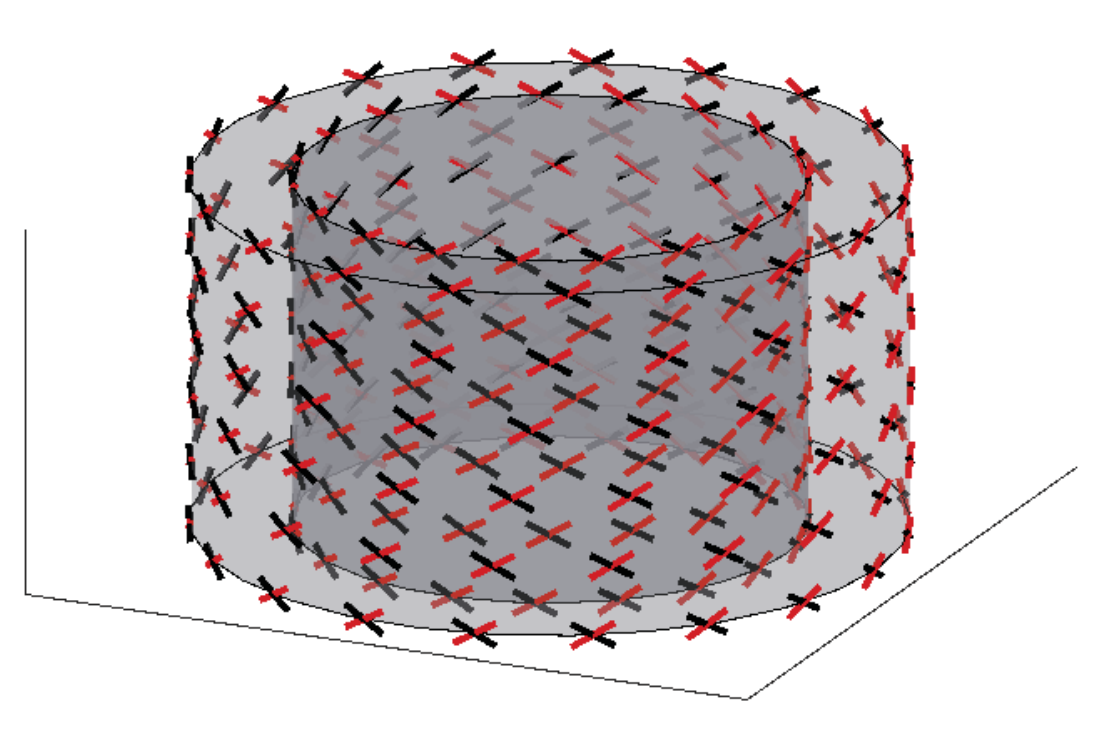}
                \caption{}
                \label{fig:gull2}
        \end{subfigure}
\end{minipage}
\begin{minipage}{.48\textwidth}
  \centering
        \begin{subfigure}[b]{\textwidth}
                \centering
                  \includegraphics[width=\textwidth ]{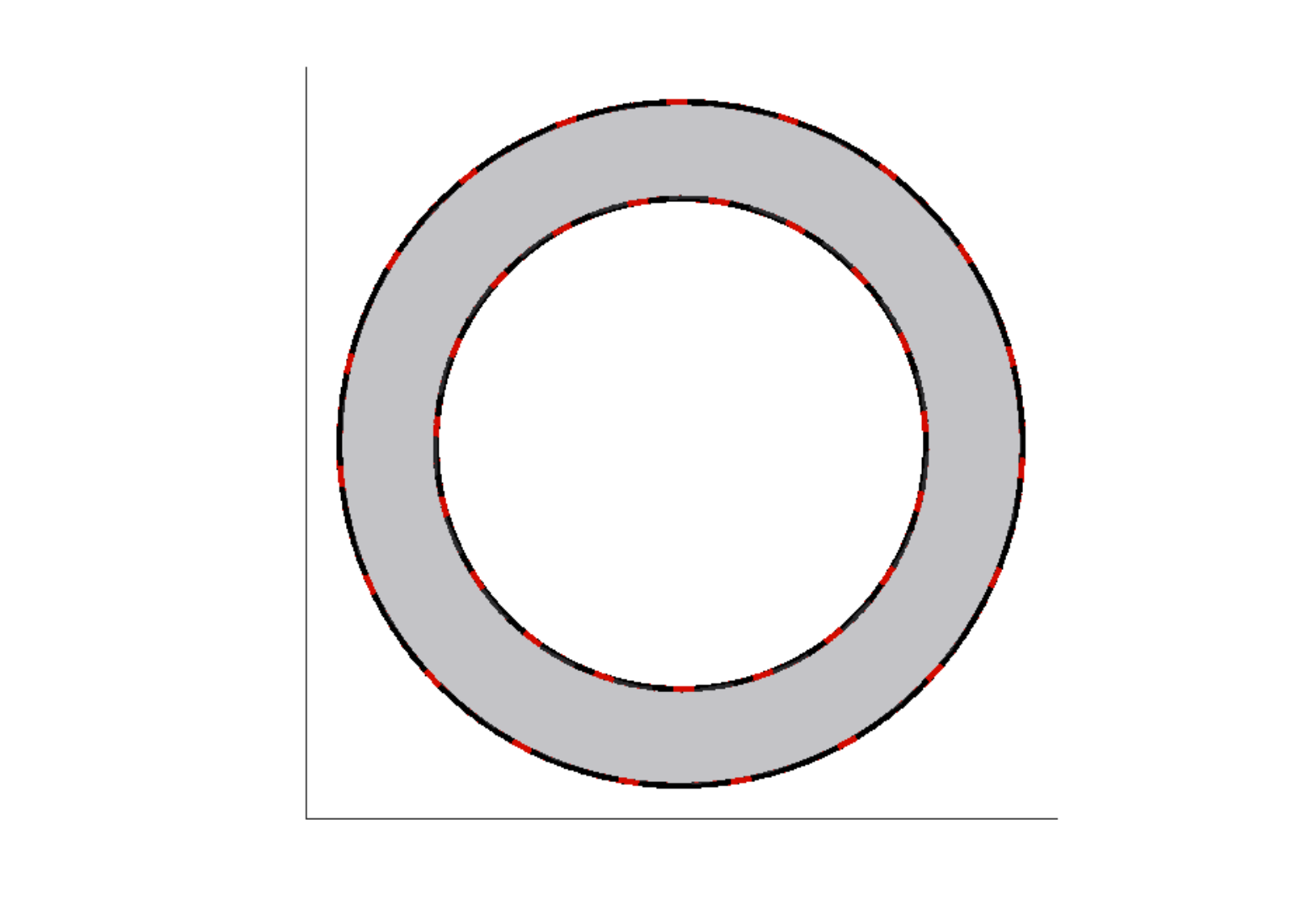}
                \caption{}
                \label{fig:gull2}
        \end{subfigure}

\end{minipage}

\begin{minipage}[t]{.99\textwidth}
  \centering
  \captionof{figure}{(a) A cylindrical two-fiber anisotropic hyperelastic solid with a helical fiber arrangement in the Lagrangian frame. (b,c) Side and top view of fibers on two sample cylindrical surfaces.}
  \label{fig:test2}
\end{minipage}
\end{figure}

As a first application, we consider a model of an arterial wall layer, which is described as a cylindrical incompressible solid along $Z=X_3$, reinforced with two families of fibers that make up helical lines around every material cylinder $X_1^2+X_2^2=R^2$. The corresponding cylindrical material coordinates will be denoted $(R, \Phi, Z)$, where $\Phi$ is the polar angle. The unit fiber direction vectors are given by
\beq\label{Eq:HelicalFiberFamily12}
\barr
\vec{A}_1(\vec{X}) &= -\cos\beta \sin\Phi \,\vec{e}_1 + \cos\beta \cos\Phi \,\vec{e}_2 + \sin\beta \,\vec{e}_3,\\[1ex]
\vec{A}_2(\vec{X}) &= -\cos\beta \sin\Phi \,\vec{e}_1 + \cos\beta \cos\Phi \,\vec{e}_2 - \sin\beta \,\vec{e}_3,
\earr
\eeq
where $\vec{e}_i$ are material Cartesian basis vectors, and $\beta$ ($0<\beta<\pi/2$) is the helical pitch angle (Figure \ref{fig:test2}). In particular, $\beta=0$ corresponds to coinciding horizontal, and $\beta=\pi/2$ to coinciding vertical fiber arrangements along the direction of $Z=X_3$. For example, in a rabbit carotid artery media layer, it was found that $\beta\simeq 29^\circ$, and in the adventitia layer of the same artery, $\beta\simeq 62^\circ$ \cite{holzapfel2000new}.

Suppose that the elastic solid occupies a cylindrical shell with inner and outer material radii $R_1$ and $R_2$ respectively:
\beq\label{eq:cyll:ann}
R_1 \leq R\leq R_2,\quad 0\leq \Phi<2\pi, \quad Z\in \mathbb{R}.
\eeq
In order to study any specific type of fully nonlinear waves propagating in such media, a deformation class \eqref{eq:Xtox} must be specified, consistent with the incompressibility assumption \eqref{eq:motion:J1}. For radially spreading shear waves with vertical displacements, the deformation class is given by
\begin{equation}\label{Eq:HyperelasticDeformationClass}
\left(\begin{array}{c}
r\\ \phi\\ z\end{array}\right) = \left(\begin{array}{c}
R\\ \Phi \\ Z + G(t,R)\end{array}\right)\text{,}
\end{equation}
where the displacement $G(t,R)$ is not assumed to be small. The deformation gradient is computed as
\begin{align}
\tens{F} &= \frac{D\left(x_1, x_2, x_3\right)}{D\left(X_1, X_2, X_3\right)} = \left(\frac{D\left(x_1, x_2, x_3\right)}{D\left(r, \phi, z\right)}\right) \left(\frac{D\left(r, \phi, z\right)}{D\left(R, \Phi, Z\right)}\right) \left(\frac{D\left(R, \Phi, Z\right)}{D\left(X_1, X_2, X_3\right)}\right) \nonumber\\[2ex]
  &= \left(\begin{array}{ccc}
1 & 0 & 0\\
0 & 1 & 0\\
\cos\Phi \pder[G(t, R)]{R} & \sin\Phi \pder[G(t, R)]{R} & 1 \end{array}\right);\label{eq:our:defclass:F}
\end{align}
it identically satisfies the incompressibility condition $J= \det {\tens{F}} = 1$.

\subsection{Nonlinear and linear radial wave models}\label{sec:nonlWgen}

Consider a general class of constitutive models of incompressible hyperelastic solids with two fiber families, with isotropic and anisotropic parts of the hyperelastic stored energy given by \eqref{eq:constit:iso} and \eqref{eq:Eaniso:2fib:gen}:
\begin{equation}\label{Eq:HyperelasticStrainEnergyDensity:gen}
W^h = U(I_1,I_2)+V(I_4, I_5, I_6, I_7, I_8).
\end{equation}
The dynamic equations for the displacement $G(t,R)$ and pressure $p(t,R)$ are derived componentwise from \eqref{eq:PK1:from:Wh}, \eqref{eq:motion:basic:lagr:mom}. The following statement is proven in Appendix \ref{app:genW}.
\begin{theorem} \label{th:genw}
Axially symmetric finite anti-plane shear displacements $G(t,R)$ of fully nonlinear incompressible hyperelastic solid reinforced with two interacting families of fibers  \eqref{Eq:HelicalFiberFamily12}, with the stored energy function \eqref{Eq:HyperelasticStrainEnergyDensity:gen},
satisfy the wave equation \eqref{G:for:genW:f}
\[
G_{tt} = \dfrac{1}{R}\dfrac{\partial }{\partial R}\left( R\, f(G_R)\right),
\]
where
\beq\label{G:for:genW:f2:f}
f(G_R)=2\, G_R (U_1+U_2+2V_5\sin^2\beta),
\eeq
and $U_i$, $V_j$ are the corresponding partial derivatives of the stored energy function \eqref{Eq:HyperelasticStrainEnergyDensity:gen}.
\end{theorem}
In particular, the incompressibility requirement \eqref{eq:motion:J1} is identically satisfied. The $Z-$component of the momentum equations \eqref{eq:motion:basic:lagr:mom} leads to the equation of motion \eqref{G:for:genW:f}.  $XY-$plane components of \eqref{eq:motion:basic:lagr:mom} yield the $R-$derivative of the hydrostatic pressure $p(t,R)$ \eqref{eq:genW:Peq}, with an additional requirement that $V$ in \eqref{Eq:HyperelasticStrainEnergyDensity:gen} satisfies the compatibility condition \eqref{eq:genW:Pcond}. The ODE \eqref{eq:genW:Peq} defines the pressure $p(t,R)$ for every solution of the wave equation \eqref{G:for:genW:f}, up to an arbitrary additive function of time.

For the deformations of the form \eqref{Eq:HyperelasticDeformationClass}, the invariants $I_j$, $j=1,..,8$, are given by \eqref{eq:genW:Ij}. It follows that for any form of the  stored energy function \eqref{Eq:HyperelasticStrainEnergyDensity:gen} where $f(G_R)$ \eqref{G:for:genW:f2:f} is a linear function of $G_R$ when computed on \eqref{eq:genW:Ij}, the wave amplitude $G(t,R)$ satisfies a \emph{linear PDE}. The pressure equation \eqref{eq:genW:Peq}, however, remains generally nonlinear.

We note that the equation of motion \eqref{G:for:genW:f} is invariant under the addition of an arbitrary function $\tilde{V}(I_4, V_6, I_8)$ to the anisotropic energy term $V$ in \eqref{Eq:HyperelasticStrainEnergyDensity:gen}.

It is straightforward to show that the wave equation \eqref{G:for:genW:f} arises from a variational principle. The action functional and the Lagrangian density for the PDE \eqref{G:for:genW:f} are given by
\[
\mathcal{S}=\displaystyle\iint_{0}^{\infty} L \, dR\, dt,\qquad L=R\left(\dfrac{G_t^2}{2}-F(G_R)\right),
\]
where $F$ is an antiderivative of $f$. Indeed, the action of the Euler operator with respect to $G$ on the Lagrangian yields
\[
\sg{E}_{G}\,L = R\left( G_{tt} -\dfrac{1}{R}\dfrac{\partial }{\partial R}( R\, f(G_R))\right)=0,
\]
which is essentially the PDE \eqref{G:for:genW:f}.


\subsection{Radial waves in Mooney-Rivlin solids with quadratic reinforcement}\label{sec:MRlin}

An important specific class of constitutive models of the hyperelastic stored energy is given by a combination of the Mooney-Rivlin and the standard (quadratic) reinforcement terms \cite{peng2010simple, cheviakov2016one}:
\begin{equation}\label{Eq:HyperelasticStrainEnergyDensity}
W^h = a\left(I_1-3\right) + b\left(I_2 - 3\right) + q_1\left(I_4-1\right)^2+q_2\left(I_6-1\right)^2+K_3 I_8^2 + K_4 I_8,
\end{equation}
involving the Mooney-Rivlin-type isotropic part with  constant parameters $a$, $b$, and an anisotropic part. Here $q_{1,2}$ are two fiber strength parameters of the corresponding two fiber families \eqref{Eq:HelicalFiberFamily12}, and $K_{3,4}$ are fiber interaction constants. The class of models \eqref{Eq:HyperelasticStrainEnergyDensity} is a subset of \eqref{Eq:HyperelasticStrainEnergyDensity:gen}; it can be viewed as a Taylor approximation of a broad class of two-fiber-family hyperelastic constitutive models
where the strain energy density $W^h$ depends on the invariants $I_1$, $I_2$, $I_4$, $I_6$, and $I_8$.

The equations of motion for the displacement $G(t,R)$ are derived componentwise from \eqref{eq:PK1:from:Wh}, \eqref{eq:motion:basic:lagr:mom}, and constitute a special case of \eqref{G:for:genW:f}, \eqref{G:for:genW:f2:f} with $U_1=a$, $U_2=b$, $V_5=0$. The following statement holds.
\begin{theorem} \label{th:linw}
For the model of a fully non-linear incompressible hyperelastic solid reinforced with two interacting fiber families \eqref{Eq:HelicalFiberFamily12}, defined by the stored energy function of the form \eqref{Eq:HyperelasticStrainEnergyDensity}, finite shear displacements in the cylinder axis direction \eqref{Eq:HyperelasticDeformationClass} propagating in the radial direction are described by solutions $G(t,R)$ of a linear wave equation \eqref{G:lin}
\[
G_{tt} = \alpha \left(G_{RR} + \dfrac{1}{R}G_R\right),\qquad \alpha = 2(a + b)=\const.
\]
\end{theorem}

Importantly, the PDE \eqref{G:lin} depends neither on the fiber parameters $q_1$, $q_2$, $K_3$, $K_4$, nor on the fiber pitch angle $\beta$, but only on the Mooney-Rivlin constants $a,b$.

For the strain energy density \eqref{Eq:HyperelasticStrainEnergyDensity}, the $R-$ and $\Phi-$projections of the momentum equations are compatible, and yield the same pressure equation
\begin{equation}\label{Eq:HHEP}
p_R=-\dfrac{\rho_0}{R}(2 b  G_R^2 +  M),
\end{equation}
which is a special case of \eqref{eq:genW:Peq}, with
\beq\label{eq:hyp:M}
M = \cos 2\beta (1+\cos 2\beta) \left(2 K_3\cos^2 2\beta + K_4\right)=\const.
\eeq

The wave  equation \eqref{G:lin} coincides with the linear wave model of small axially symmetric vertical oscillations of an elastic membrane, but is obtained here without any assumption on the smallness of the displacement $G$ or any other parameters. Interestingly, the condition \eqref{Eq:HHEP} defining the pressure also does not involve the fiber stretch constants $q_1$, $q_2$, but does depend on the fiber interaction constants $K_3$, $K_4$, and the fiber pitch angle $\beta$, or specifically, the angle $2\beta$ between the fiber family directions.

\subsection{Initial-boundary value problems}

\subsubsection{Dirichlet and Neumann problems in an annulus}

A well-posed initial-boundary value problem (IBVP) for the radial s-wave equation \eqref{G:for:genW:f} consists of two boundary and two initial conditions, and is naturally stated in a cylindrical annulus $R\in [R_1,R_2]$, $0<R_1<R_2$. For example, a Dirichlet problem corresponding to the inner arterial wall being periodically driven according to $g(t)$ in the vertical direction (e.g., by a blood flow pulsation), and the right boundary being fixed, involves boundary conditions
\beq\label{eq:linw:Diri}
G(R_1,t) = g(t), \quad G(R_2,t) = 0
\eeq
stated for $t>0$. Perhaps a more realistic yet elementary model of the action of blood flow on the inner wall of the blood vessel can be described by Neumann boundary conditions corresponding to a periodic vertical traction forcing at the inner wall $R=R_1$, and a free outer boundary $R=R_2$:
\beq\label{eq:linw:Neu}
G_R(R_1, t) = g(t),\quad G_R(R_2,t) = 0,
\eeq
with a traction force prescribed by $g(t)$, $t>0$. A set of boundary conditions \eqref{eq:linw:Diri} or \eqref{eq:linw:Neu} is supplemented with initial conditions
\beq\label{eq:linw:IC}
G(R,0) = G_0(R),\quad G_t(R,0) = G_1(R),\quad R_1\leq R\leq R_2.
\eeq

When the PDE \eqref{G:for:genW:f} is linear, for example, in the case \eqref{G:lin}, the Dirichlet IBVP \eqref{G:lin}, \eqref{eq:linw:Diri}, \eqref{eq:linw:IC} and the Neumann IBVP
\eqref{G:lin}, \eqref{eq:linw:Neu}, \eqref{eq:linw:IC} are solved by separation of variables; examples of explicit solutions for zero initial conditions $G_0(R)=G_1(R)=0$ are given in Appendix \ref{app:solDN}.

\subsubsection{The boundary value problem for a two-layer medium}

In the modeling of arteries, it is important to take into account their multi-layered structure, in particular, the most mechanically significant layers are the \emph{adventitia} (outer layer) and the \emph{media} (inner layer) \cite{holzapfel2000new, gasser2006hyperelastic}. Both of these layers contain helical collagen fibers, with different pitch angles.

\begin{figure}[htbp]
\begin{center}
\includegraphics[width=0.4\textwidth]{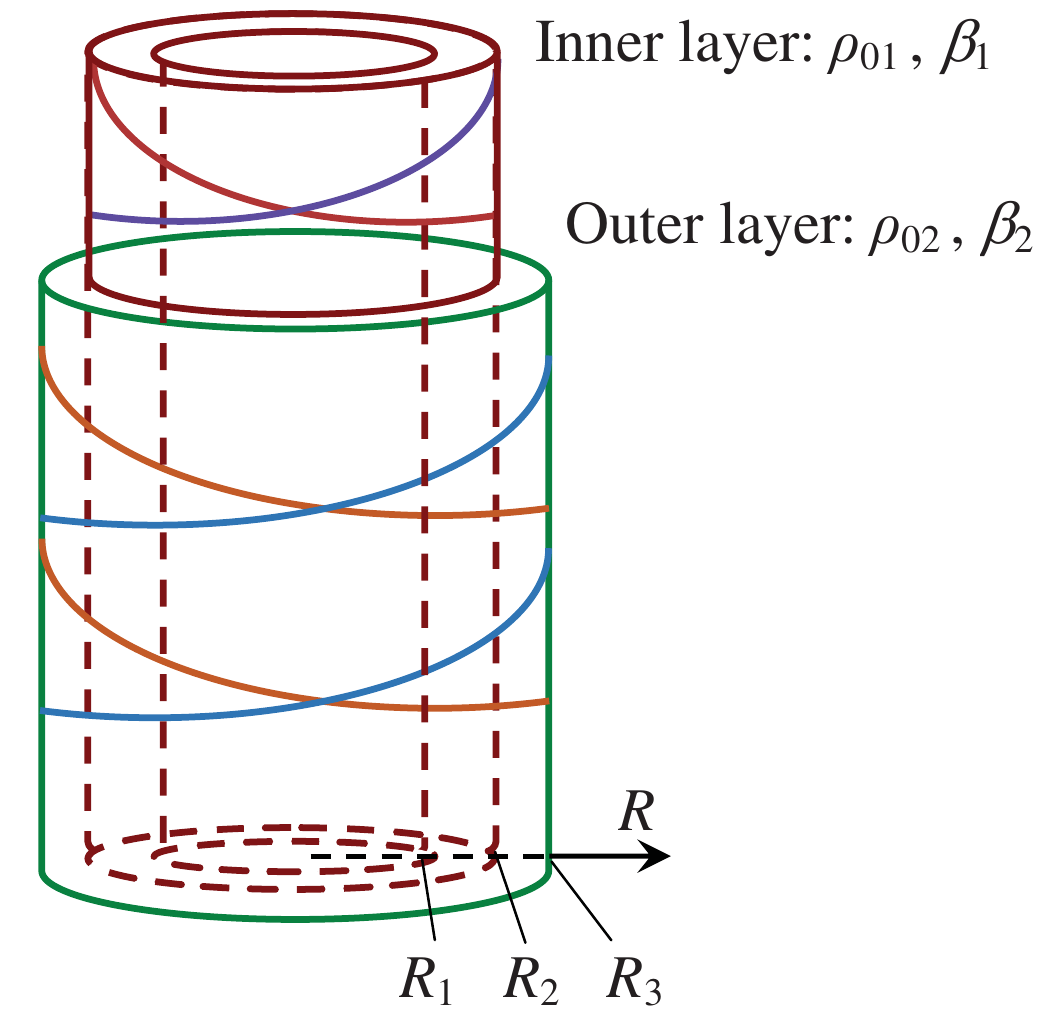}
\end{center}
  \caption{A two-layer cylindrical solid with two families of helical fibers in each layer.
  \label{fig:2materials}
  }
\end{figure}

Consider a two-layer cylindrical solid, with the inner layer occupying the annulus $R_1 \leq R\leq R_2$, and the outer layer $R_2 \leq R\leq R_3$ (Figure \ref{fig:2materials}), modeled as a hyperelastic medium with a general stored energy function \eqref{Eq:HyperelasticStrainEnergyDensity:gen}. An IBVP for the PDE \eqref{G:for:genW:f}  with general initial conditions and general linear (Robin) boundary conditions is given by
\beq\label{eq:linw:2layer}
\barr
G_{tt}= \dfrac{1}{R}\dfrac{\partial }{\partial R}\left( R\, f(G_R)\right),\qquad
f(G_R)=\left\{\barr  f_1(G_R),~~R_1 < R< R_2,\\ f_2(G_R),~~R_2 < R< R_3;\earr\right.\\[2ex]
G(R,0) = G_0(R),\quad G_t(R,0) = G_1(R),\quad R_1 < R< R_3;\\[1ex]
q_1 G(R_1, t)+q_2 G_R(R_1, t) =  g_1(t),\\[1ex]
q_3 G(R_2, t)+q_4 G_R(R_3, t) =  g_2(t),
\earr
\eeq
where $f_1$ and $f_2$ are respective response functions for the inner and the outer layer. In particular, the fiber pitch angles $\beta_1$, $\beta_2$ for the two layers may be different; for example, they were measured to equal approximately $29^\circ$ and $62^\circ$ respectively for the \emph{media} and \emph{adventitia} layers of the carotid artery of a rabbit \cite{holzapfel2000new}. In \eqref{eq:linw:2layer}, the functions $G_0(R)$ and $G_1(R)$ determine the initial conditions; the constants and $q_i$, $i=1,\ldots,4$ and the functions $g_1(t)$ and $g_2(t)$ determine the type and forcing terms of the boundary conditions.

In the case of a Mooney-Rivlin/quadratic reinforcement stored energy \eqref{Eq:HyperelasticStrainEnergyDensity}, the PDE \eqref{G:lin} is linear, the fiber effects will not affect the shear material displacements, and the coefficient $\alpha=\alpha(R)$ in \eqref{G:lin} is piecewise-constant, equal to $\alpha_1=2(a_1 + b_1)$ for $R_1 < R< R_2$, and  $\alpha_2=2(a_2 + b_2)$ for $R_2 < R< R_3$, with $a_1, b_1$ and $a_2, b_2$ denoting Mooney-Rivlin parameters for the inner and the outer layer.

The two-layer model must also include physical contact conditions at $R=R_2$, which for the linear wave model \eqref{G:lin} take the form
\beq\label{eq:linw:2layer:cond}
\barr
G(R_2-0,t)=G(R_2+0,t),\\[1ex]
\rho_{01}\alpha_1 G_R (R_2-0,t)=\rho_{02}\alpha_2 G_R (R_2+0,t).
\earr
\eeq
Here the first equation ensures the continuity of the displacement, and the second one expresses the continuity requirement of the shear ($Z-$directed) component of the traction force $\vec{T}=\tens{P}\vec{N}$ acting on a unit area of the cylinder boundary (i.e., the third law of Newton). Here $\rho_{01}=\rho_0(R_2-0)$ and $\rho_{02}=\rho_0(R_2+0)$ respectively denote the densities of the inner and outer layers on the material interface $R=R_2$.

\section{A Modified Fiber Model}\label{sec:4:modif:fibers}

\begin{figure}[htbp]
\centering
        \begin{subfigure}[b]{0.44\textwidth}
                \centering
                \includegraphics[width=\textwidth]{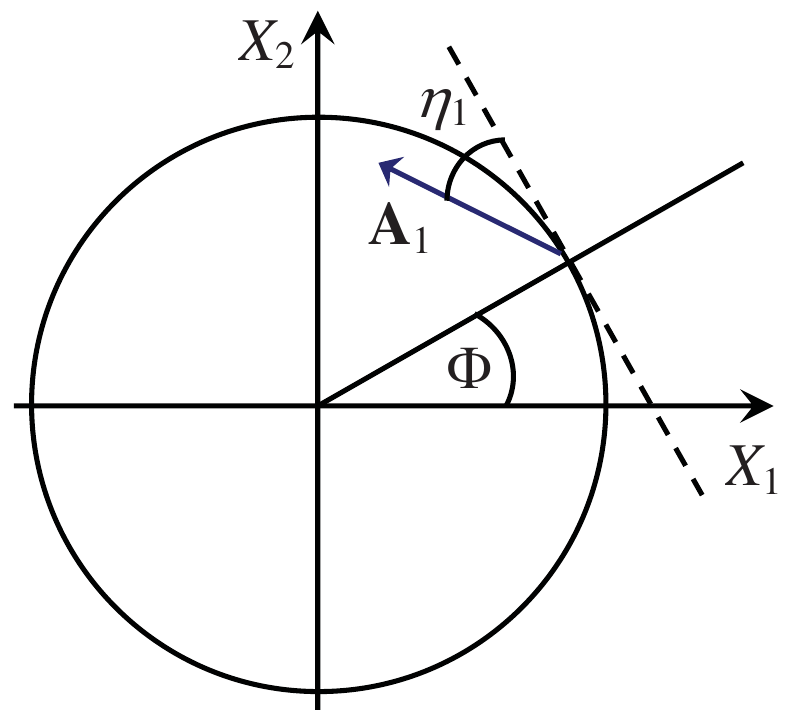}
                \caption{}
                \label{fig:gull}
        \end{subfigure}%

\medskip

\begin{minipage}{.48\textwidth}
  \centering
        \begin{subfigure}[b]{\textwidth}
                \centering
                  \includegraphics[width=\textwidth]{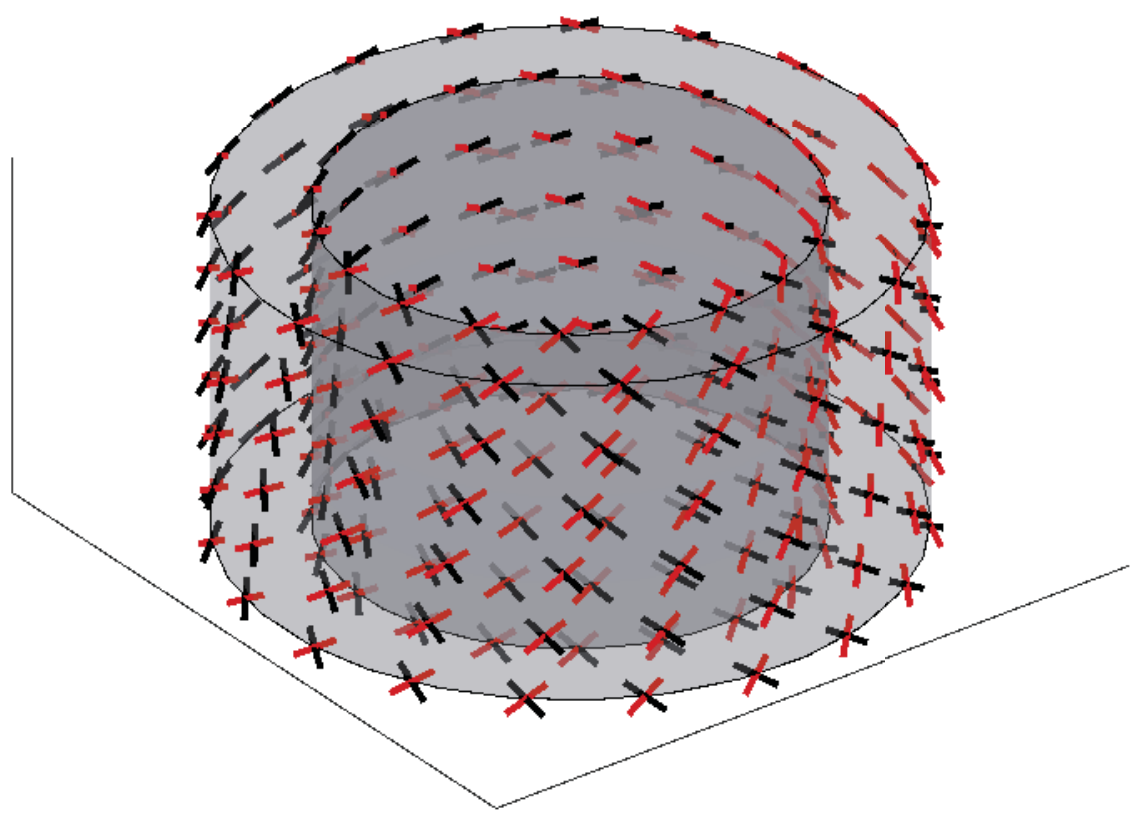}
                \caption{}
                \label{fig:gull2}
        \end{subfigure}
\end{minipage}
\begin{minipage}{.48\textwidth}
  \centering
        \begin{subfigure}[b]{\textwidth}
                \centering
                  \includegraphics[width=0.8 \textwidth ]{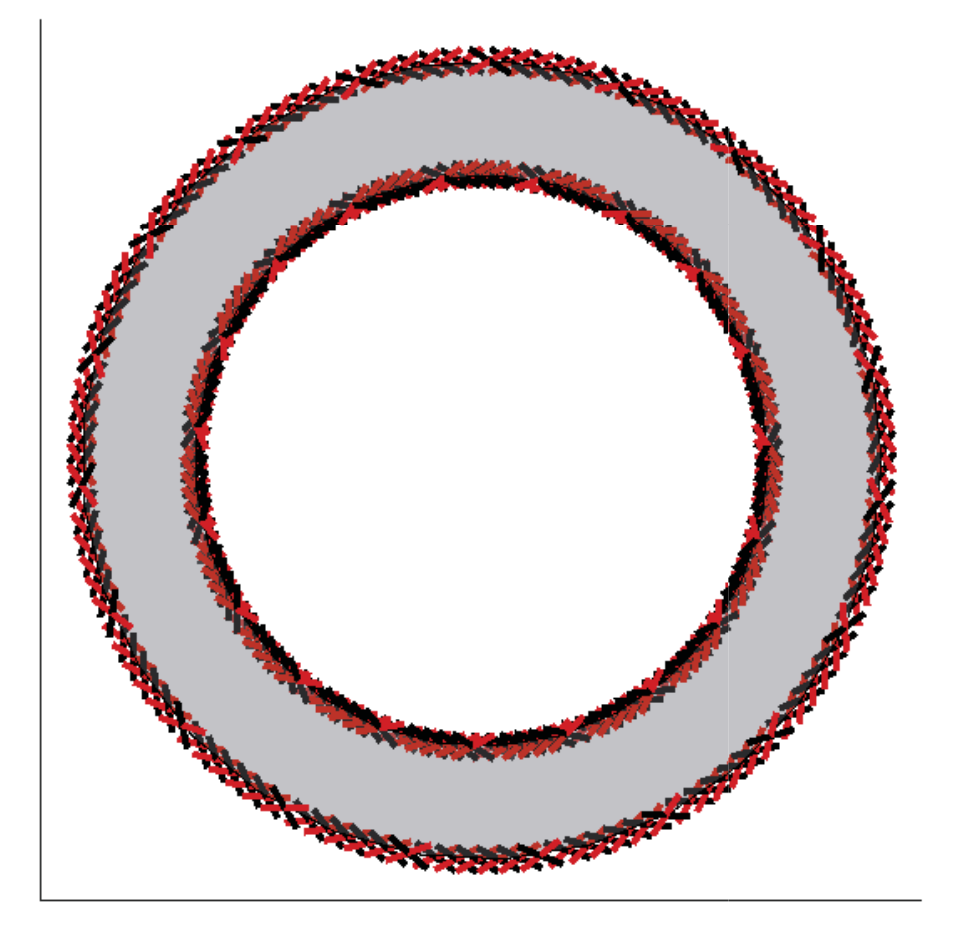}
                \caption{}
                \label{fig:gull2}
        \end{subfigure}

\end{minipage}

\begin{minipage}[t]{.99\textwidth}
  \centering
  \captionof{figure}{(a) Projection on the horizontal plane of a fiber from the family $\vec{A}_1$ in the modified fiber model. (b,c) Side and top view of two fiber families on two sample cylindrical surfaces for the modified fiber model with $\eta_1=\pi/6$, $\eta_2=-\pi/6$.}
  \label{fig:modfib}
\end{minipage}
\end{figure}

As seen in Section \ref{sec:MRlin}, for an incompressible medium with fiber families tangent to nested cylinders, subject to the Mooney-Rivlin constitutive relation with the quadratic fiber reinforcement terms \eqref{Eq:HyperelasticStrainEnergyDensity}, as well as for some more general constitutive relations (see Section \ref{sec:nonlWgen}),
fibers have no effect on the propagation of radial shear waves of the form \eqref{Eq:HyperelasticDeformationClass}. In particular, material displacements for such waves are described by the linear wave equation \eqref{G:lin}. We now extend the model of Section \ref{Sec:HyperelasticLinearModel}, allowing both fiber families to have nonzero projections on the radial direction. For each fiber family, a nonzero angle $\eta$ corresponds to a nonzero radial fiber projection (see Figure \ref{fig:modfib}a).

Without loss of generality, it is convenient to write the radial projection parameters $\eta_{1,2}$ for the two fiber families as
\beq\label{eq:mod:etas:delta}
\eta_1 = \eta + \delta, \quad \eta_2 = \eta - \delta.
\eeq
The unit fiber direction vectors for the two fiber families are consequently given by (cf. \eqref{Eq:HelicalFiberFamily12})
\beq\label{Eq:ModFiberFamily12}
\barr
\vec{A_1} &= -\cos\beta\sin{(\Phi + \eta + \delta)}\,\vec{e_1} + \cos\beta\cos{(\Phi + \eta + \delta)}\, \vec{e_2} + \sin\beta\, \vec{e_3},\\[1ex]
\vec{A_2} &= -\cos\beta\sin{(\Phi + \eta - \delta)}\,\vec{e_1} + \cos\beta\cos{(\Phi + \eta - \delta)}\, \vec{e_2} - \sin\beta \,\vec{e_3},
\earr
\eeq
where $\vec{e}_i$ are material Cartesian basis vectors, $\Phi$ is the cylindrical polar angle, and $\beta$ is the fiber helical pitch angle (Figure \ref{fig:test2}a).

%

Using these generalized fiber family orientations, we repeat the steps of Section \ref{Sec:HyperelasticLinearModel} to derive the equations of motion for the fully nonlinear waves moving in the radial direction, with displacements determined by the deformation class \eqref{Eq:HyperelasticDeformationClass}. The underlying constitutive model is the same as that of Section \ref{Sec:HyperelasticLinearModel}, with the Mooney-Rivlin/quadratic reinforcement stored energy function \eqref{Eq:HyperelasticStrainEnergyDensity}.

As before, for the current setup, the incompressibility condition \eqref{eq:motion:J1} is identically satisfied. The three momentum equations \eqref{eq:motion:basic:lagr:mom} reduce to three PDEs involving the unknown vertical displacements $G(t,R)$ and the hydrostatic pressure $p(t,R)$. The equations of motion involve nine arbitrary material parameters
\beq\label{eq:mod:all:params}
a, b,  q_1, q_2, K_3, K_4, \beta, \eta, \delta,
\eeq
specifically, the Mooney-Rivlin isotropic elasticity parameters $a,b$, the anisotropic fiber strength/interaction parameters $q_1, q_2, K_3, K_4$, and the fiber orientation angles $\beta, \eta, \delta$.

While, as before, the $Z-$projection of the momentum equations \eqref{eq:motion:basic:lagr:mom} yields the wave equation on the vertical displacements $G(t,R)$, the PDEs obtained for the $X-$ and $Y-$projections of the momentum equations contain the pressure equation and a compatibility condition on the displacements $G(t,R)$, which was not present in an earlier model of Section \ref{Sec:HyperelasticLinearModel}, and is due to a nonzero fiber radial projection. For some classes of materials, the compatibility condition vanishes. The following statement is proven by direct computation.

\begin{theorem}
Radial shear waves in an incompressible hyperelastic medium governed by Mooney-Rivlin/quadratic reinforcement constitutive relation \eqref{Eq:HyperelasticStrainEnergyDensity}, where the two fiber families \eqref{Eq:ModFiberFamily12} have symmetric radial projection angles $\eta_1 = - \eta_2=\delta$ and equal fiber strengths $q_1=q_2$, are described by vertical displacements $G(t,R)$ satisfying a nonlinear wave equation
\begin{equation}\label{Eq:MHE3}
G_{tt} = N_1\left(G_{RR} + \dfrac{1}{R}G_R\right) + N_2 G_R\left(2G_{RR} + \dfrac{1}{R}G_R\right) + N_3 G_R^2 \left(3G_{RR} + \dfrac{1}{R}G_R\right) + \dfrac{N_4}{R}\, ,
\end{equation}
where the constant coefficients $N_i$  are material parameters given by
\begin{equation}\label{Eq:ModifiedHyperbolicSimplifiedCoefficients}
\barr
N_1 &= \alpha - 2\cos^2\beta\sin^2\delta\left[2 K_3 \left(2 \cos^2\beta(1+\cos^2\delta) - 3\right)(2 \cos^2\beta \cos^2\delta - 1)^2\right.\\
& \left. + K_4\left(2 \cos^2\beta \cos^2\delta - 1\right) - 8 q\sin^2\beta\right]\, ,\\[2ex]
N_2 &= -12 \sin\beta \cos^3\beta \sin^3\delta \left(K_3\left(2\cos^2\beta \cos^2\delta - 1\right)^2 + 2q\right)\, ,\\[2ex]
N_3 &= 4 \cos^4\beta \sin^4\delta\left(K_3 \left(2\cos^2\beta \cos^2\delta - 1\right)^2 + 2q\right)\, ,\\[2ex]
N_4 &= 2 \sin\beta \cos\beta \sin\delta\left(2\cos^2\beta\cos^2\delta-1\right) \left(2K_3\left(2\cos^2\beta\cos^2\delta-1\right)^2 + K_4\right)\, ,
\earr
\end{equation}
and $\alpha = 2(a+b)$.
\end{theorem}

The wave equation \eqref{Eq:MHE3} describes radial waves in a medium where one helical fiber family is turned ``inwards", and the other ``outwards", by the same angle $\delta$ (see Figure \ref{fig:modfib}b,c). The nonlinear PDE \eqref{Eq:MHE3} is thus an extension of the linear PDE \eqref{G:lin} in Theorem \ref{th:linw} onto the case of fibers with nonzero projection on the wave propagation direction. It is consistent with the model of Section \ref{Sec:HyperelasticLinearModel}, reducing to the linear PDE \eqref{G:lin} when $\delta=0$. When $\delta$ is a small parameter, $|\delta|\ll 1$, it is straightforward to compute the leading terms for the coefficients $N_i$ \eqref{Eq:ModifiedHyperbolicSimplifiedCoefficients}:
\begin{equation}\label{Eq:ModifiedHyperbolicSimplifiedCoefficients:smallD}
\barr
N_1 &= \alpha - 2\cos^2\beta \big(2 K_3 \cos^22 \beta(4 \cos^2\beta - 3) + K_4 \cos2 \beta - 8q\sin^2\beta\big)\delta^2+\mathcal{O}(\delta^4),\\[2ex]
N_2 &= - 12 \sin\beta \cos^3\beta \left(K_3 \cos^2 2 \beta + 2 q\right)\delta^3+\mathcal{O}(\delta^5) ,\\[2ex]
N_3 &= 4 \cos^4\beta \left(K_3 \cos^2 2 \beta + 2 q\right)\delta^4+\mathcal{O}(\delta^6) ,\\[2ex]
N_4 &= 2 \sin \beta  \cos \beta  \cos 2 \beta \left(2 K_3 \cos^2 2 \beta  + K_4\right)\delta+\mathcal{O}(\delta^3).
\earr
\end{equation}
This yields the term ordering for perturbation theory analysis of the PDEs \eqref{Eq:MHE3} as $\delta\to 0$.

\medskip

The nonlinear PDEs \eqref{Eq:MHE3} can be rewritten in a divergence form \eqref{G:4N}, and thus belong to the family of equations \eqref{G:for:genW:f} with
\[
f(G_R)=N_1G_{R}+  N_2 G_R^2 + N_3 G_R^3 + N_4.
\]
Yet the equations \eqref{Eq:MHE3}, \eqref{G:4N} were obtained under different physical assumptions, specifically, a different arrangement of elastic fiber families.

%


\subsection{The horizontal fiber model and a numerical example}

A particularly simple case of the nonlinear wave equations \eqref{Eq:MHE3} or \eqref{G:4N} can be obtained by taking the pitch angle $\beta=0$. This case corresponds to an ``almost-circular" arrangement of the two fiber families, and one has $N_2=N_4=0$. The wave equation \eqref{Eq:MHE3} then assumes a simpler form,
differing from the linear wave equation \eqref{G:lin} by a single nonlinear term. Let $R_c$ be some characteristic radius (e.g., the outer radius $R_2$ of the annular domain). Then in terms of the dimensionless (starred) variables
\begin{align*}
G = R_c G^{*}\, \quad R = R_c R^{*}, \quad t=\frac{R_c}{\sqrt{N_1}} t^{*},
\end{align*}
the PDE \eqref{Eq:MHE3} becomes, after dropping the asterisks,
\begin{equation}\label{Eq:ModifiedFiberNondimPDE}
G_{tt} = G_{RR} + \dfrac{G_{R}}{R} + c G_{R}^2\left(3 G_{RR} + \dfrac{G_{R}}{R}\right) = \dfrac{1}{R} \dfrac{\partial}{\partial R}  \left[ RG_{R}\left( 1 +  c G_R^2 \right)\right]\,.
\end{equation}
In \eqref{Eq:ModifiedFiberNondimPDE}, $c = {N_3}/{N_1}$ is the parameter controlling the nonlinearity; its leading-order expansion as  $\delta \to 0$ is given by
\[
c = \dfrac{4}{\alpha}\left(K_3 + 2 q\right)\, \delta^4 + \mathcal{O}(\delta^6),
\]
typically small for small $|\delta|$, and inversely proportional to the Mooney-Rivlin parameter $\alpha$.



%
%
%
%
%
%
%

\medskip

\begin{figure}[htbp]
  	\begin{center}
		\includegraphics[ width = 0.8\textwidth]{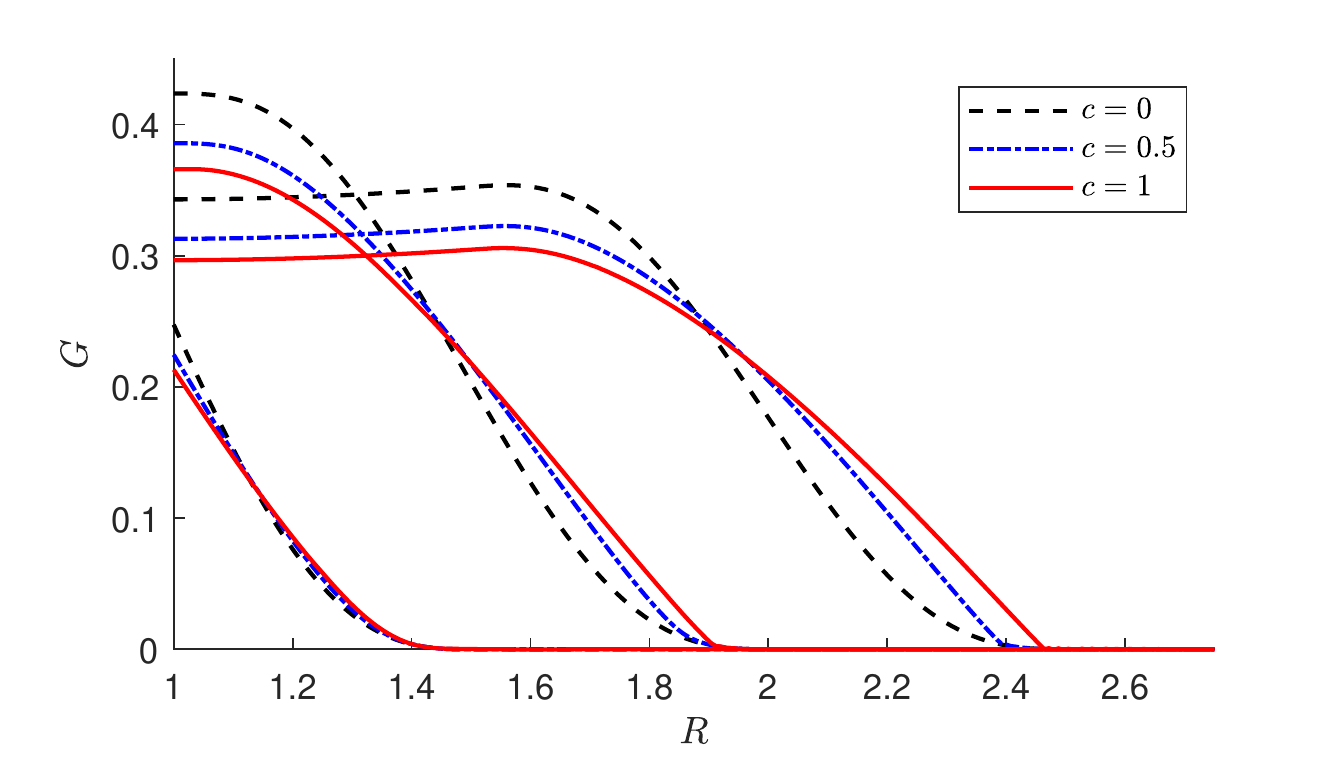}
  	\end{center}
    \caption{\label{fig:NumericResults} Numerical solutions of the PDE \eqref{Eq:ModifiedFiberNondimPDE} for  zero initial conditions and boundary conditions \eqref{eq:linw:Neu}, \eqref{eq:linw:Neugt} in the linear case $c=0$ (black dashed curve) and nonlinear cases $c=0.5$ (blue dot-dash curve) and $c=1$ (solid red curve) at the dimensionless time values $t=0.5$, $1$, and $1.5$, left to right. (Color online.)}
\end{figure}

In order to compare the original and the modified helical fiber models, we numerically simulate the dimensionless PDE \eqref{Eq:ModifiedFiberNondimPDE} for various values of $c$. The simulation is conducted in the dimensionless space-time domain $R_1 \leq R\leq R_2$, $R_1=1$, $R_2=3$, for zero initial conditions, and Neumann boundary conditions \eqref{eq:linw:Neu} with a localized nonnegative dimensionless boundary forcing
\beq\label{eq:linw:Neugt}
g(t)=G(R_1,t)=\left\{\barr  16 t^2 (t-1)^2,& 0 \leq t\leq 1,\\ 0,& t>1,\earr\right.
\eeq
with the unit $C$-norm, applied to the inner cylindrical wall. Such a setup provides a primitive model of, for example, the shear stress on the arterial wall caused by blood flow within a single heartbeat. The plots of the corresponding solutions of the PDE \eqref{Eq:ModifiedFiberNondimPDE} in the linear case $c=0$ and nonlinear cases $c=0.5, 1$ are shown in Figure \ref{fig:NumericResults}. In particular, it is observed that in the nonlinear cases, for $t>0$, the solutions develop a corner-type singularity (a jump in the derivative $G_{R}$) when $G=0$. [The numerical solutions here and below are computed using COMSOL Multiphysics finite element solver.]

%

\section{A Viscoelastic Model} \label{sec:visc}




Another nonlinear extension of the radial shear wave propagation model in a medium
with helical fibers described in Section \ref{Sec:HyperelasticLinearModel} may be obtained by taking into account viscoelastic effects (Section \ref{sec:visco:framew}). As a simple example, following Ref.~\cite{cheviakov2016one}, we let the viscoelastic strain energy component take the form
\begin{equation}\label{eq:W:visc:ours}
W^v = \frac{\mu_1}{4} J_2 \left(I_1-3\right) + \frac{\mu_2}{2} J_{9,1} \left(I_4-1\right)^2 + \frac{\mu_3}{2}J_{9,2}\left(I_6 -1\right)^2\, ,
\end{equation}
with viscosity parameters $\mu_i$, $i=1,2,3$. The viscoelastic potential \eqref{eq:W:visc:ours} corresponds, for example, to the leading Taylor terms of the viscoelastic potential of Pioletti and Rakotomanana \cite{pioletti2000non} (see our formula \eqref{eq:MerG:Wv} in Section \ref{sec:visco:framew}), adapted to include two fiber families.

We consider the same shear wave-type deformation class \eqref{Eq:HyperelasticDeformationClass} as in Section \ref{Sec:HyperelasticLinearModel}, with two identical helical fiber families ($\mu_2=\mu_3$) with orientations given by \eqref{Eq:HelicalFiberFamily12}, and the same hyperelastic energy component $W_h$ \eqref{Eq:HyperelasticStrainEnergyDensity}. We again are interested in the situation of the constant material density ($\rho_0=\const$) and no external forces ($\vec{Q}=0$). Computing the modified viscoelastic form \eqref{eq:PK2:viscoel:tot} of the second Piola-Kirchhoff tensor formula, we obtain the first Piola-Kirchhoff stress tensor $\tens{P} = \tens{F}\,\tens{S}$. The equations of motion \eqref{eq:motion:basic:lagr:mom} in the $XY-$ plane consequently lead to the pressure equation
\beq \label{Eq:VE:peq}
p_R = -\dfrac{\rho_0}{R}\Big(2G_R^2 b + M - 2\mu_1 G_R^2\left[G_R \left(G_{tR}+R G_{tRR} \right) + 3 R G_{tR}G_{RR}\right]\Big)\, ,
\eeq
where $M$ as the same constant as in \eqref{eq:hyp:M}. The $Z$-component of the equations of motion yields a nonlinear PDE
\beq \label{Eq:VE:Geq}
G_{tt} = \alpha\left(\dfrac{G_R}{R}+G_{RR}\right) + \mu_1 G_R\left[G_R\left(\dfrac{G_{tR}}{R}+G_{tRR} \right)\left(1+2 G_R^2\right) + 2G_{tR}G_{RR}\left(1+4 G_R^2\right)\right].
\eeq
One can show that the PDE \eqref{Eq:VE:Geq} can be written in an equivalent,  more compact divergence form \eqref{Geq:visco}. The following result has been established.
\begin{theorem} \label{th:viscPR3}
For the model of a fully nonlinear incompressible hyper-viscoelastic solid reinforced with two interacting helical fiber families \eqref{Eq:HelicalFiberFamily12}, defined by the hyperelastic stored energy function \eqref{Eq:HyperelasticStrainEnergyDensity}
and the viscoelastic potential \eqref{eq:W:visc:ours}, finite shear displacements in the direction of the cylinder axis \eqref{Eq:HyperelasticDeformationClass}, propagating in the radial direction, are described by solutions $G(t,R)$ of a nonlinear PDE \eqref{Geq:visco}. The latter PDE involves neither of the material fiber parameters $q_1$, $q_2$, $K_3$, $K_4$, $\mu_2$ nor the fiber helical pitch angle $\beta$, but depends only on the Mooney-Rivlin constants $a,b$ and the principal isotropic viscosity coefficient $\mu_1$.
\end{theorem}


Similarly to the linear case, the hydrostatic pressure $p(t,R)$ is determined by its $R$-derivative \eqref{Eq:VE:peq} in terms of the displacement $G(t,R)$; it depends on the fiber interaction parameters $K_3$ and $K_4$ through \eqref{eq:hyp:M}.

The model of fully nonlinear displacements described by the PDEs \eqref{Eq:VE:peq} and \eqref{Eq:VE:Geq} generalizes the linear hyperelastic model \eqref{G:lin}, \eqref{Eq:HHEP}. It admits a scaling equivalence transformation, allowing for a change of variables
\begin{align*}
G = \dfrac{G^*}{\sqrt{\alpha}}, \quad R = \dfrac{R^*}{\sqrt{\alpha}}, \quad t=\frac{t^{*}}{\alpha} ,
\end{align*}
that maps the PDE \eqref{Eq:VE:Geq} (or \eqref{Geq:visco}) into a simpler form with $\alpha^*=1$. Setting $\mu_1^*=\mu_1=\mu$ and dropping the asterisks, we obtain an equivalent dimensionless PDE
\beq \label{Geq:visco:star}
G_{tt} = \dfrac{1}{R}\dfrac{\partial}{\partial R}\Big(R G_R\left[1+\mu G_R G_{tR}\left(1+2 G_R^2\right)\right]\Big)\,.
\eeq
involving a single parameter.

\begin{figure}[htbp]
  	\begin{center}
		\includegraphics[ width = 0.8\textwidth]{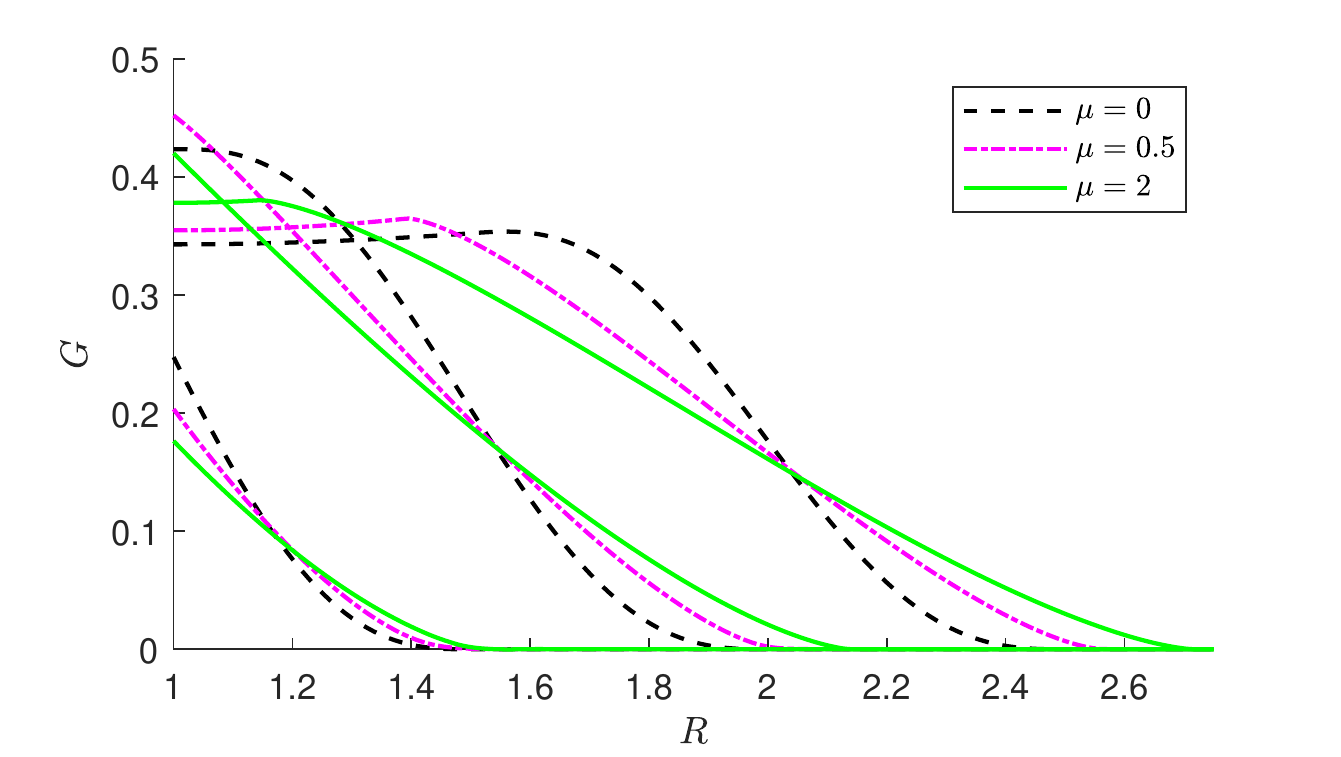}
  	\end{center}
    \caption{\label{fig:NumericVisc} Numerical solutions of the dimensionless viscoelastic PDE model  \eqref{Geq:visco:star} with zero initial conditions and Neumann boundary conditions \eqref{eq:linw:Neu}, \eqref{eq:linw:Neugt} in the linear case $\mu=0$ (black dashed curve) and nonlinear cases $\mu=0.5$ (purple dot-dash curve) and $\mu=2$ (solid green curve) at the dimensionless time values $t=0.5$, $1$, and $1.5$, left to right. (Color online.)}
\end{figure}

The nonlinear viscoelastic term in \eqref{Geq:visco:star} containing mixed space-time derivatives are responsible for diffusion-type effects. As an example illustrating these effects, we consider an initial-boundary  value problem consisting of the dimensionless PDE \eqref{Geq:visco:star}, zero initial conditions, and Neumann boundary conditons \eqref{eq:linw:Neu} with the time-localized forcing \eqref{eq:linw:Neugt} on the inner cylindrical boundary.  Sample plots of the solutions of the PDE \eqref{Geq:visco:star} in the linear case $\mu_1=0$ and nonlinear cases $\mu=0.5, 2$ are given  in Figure \ref{fig:NumericVisc}.

\begin{figure}[htbp]
  	\begin{center}
		\includegraphics[ width = 0.8\textwidth]{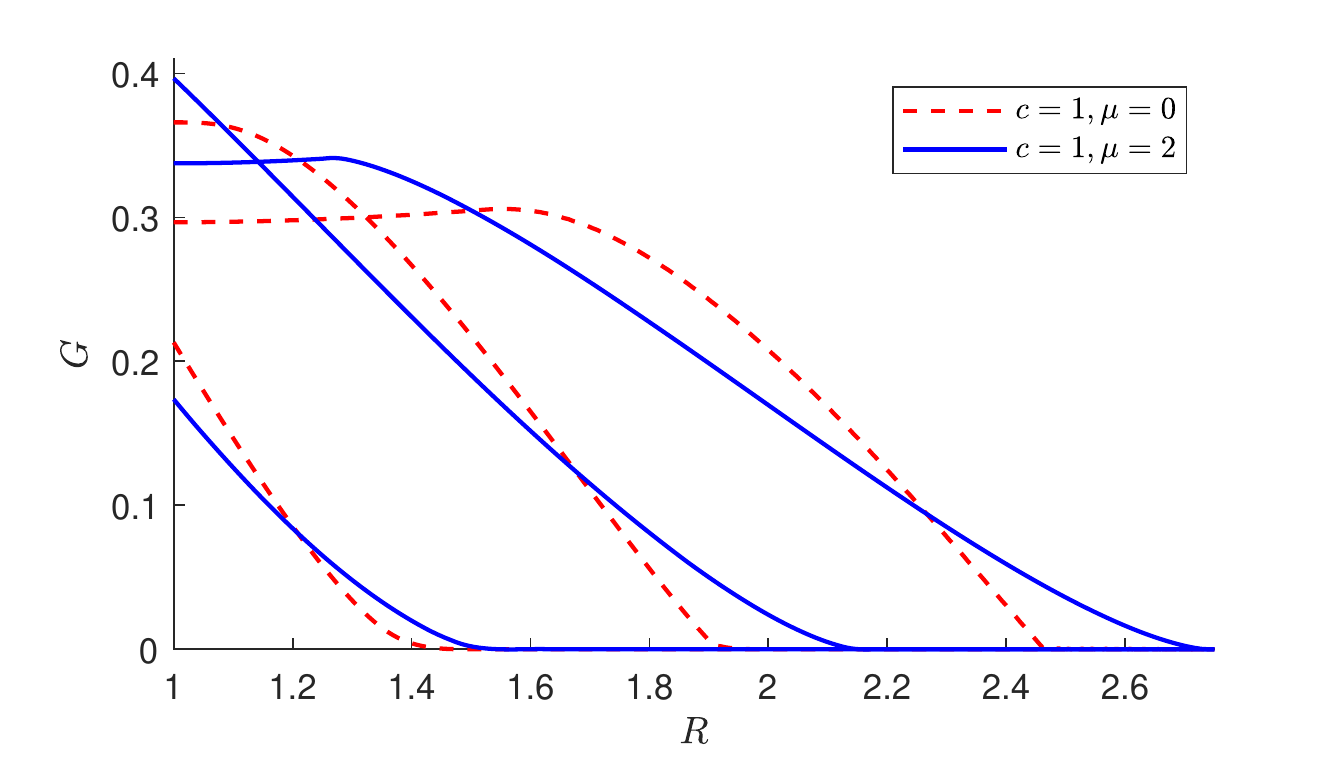}
  	\end{center}
    \caption{\label{fig:NumericViscCstar} Numerical solutions of the combined PDE \eqref{Geq:Cs:visco:star} with zero initial conditions and Neumann  boundary conditions \eqref{eq:linw:Neu}, \eqref{eq:linw:Neugt}: comparison of the inviscid case $c=1$, $\mu=0$ (red dashed curve) and the viscous case $c=1$, $\mu=2$ (blue solid curve) for the time values $t=0.5$, $1$, and $1.5$, left to right. (Color online.)}
\end{figure}

It is of further interest to compare the effects of both nonlinear terms related to modified fibers (dimensionless PDE \eqref{Eq:ModifiedFiberNondimPDE}) and viscosity (dimensionless PDE \eqref{Geq:visco:star}). Forming a joint dimensionless equation
\beq \label{Geq:Cs:visco:star}
G_{tt} = \dfrac{1}{R}\dfrac{\partial}{\partial R}\Big(R G_R\left[1+c G_R^2 +\mu G_R G_{tR}\left(1+2 G_R^2\right)\right]\Big)\,
\eeq
and solving the same Neumann problem with boundary conditions \eqref{eq:linw:Neu}, \eqref{eq:linw:Neugt} for $c=1$ and $\mu=2$, we compare it to the case $\mu=0$ (no viscosity, also shown by the solid red curve in Figure \ref{fig:NumericResults}). The result indicates that viscosity effects lead to diffusion-type damping in the solutions, and prevent the formation of the singularity at $G=0$ (Figure \ref{fig:NumericViscCstar}).

%
%
%
%



\section{Discussion}\label{sec:Discussion}

In the current paper, the family of scalar nonlinear wave equations \eqref{G:for:genW:f}, \eqref{G:for:genW:f2:f} was derived, describing finite displacements $G(t,R)$ in fully nonlinear radial shear waves propagating in a hyperelastic incompressible solid reinforced by two families of helical fibers, for a general form of a stored energy function \eqref{Eq:HyperelasticStrainEnergyDensity:gen}. It was shown that for a broad class of constitutive relations \eqref{Eq:HyperelasticStrainEnergyDensity:gen}, in particular, Mooney-Rivlin solids with standard (quadratic) reinforcement \eqref{Eq:HyperelasticStrainEnergyDensity}, such waves are described by a linear PDE \eqref{G:lin}. Initial-boundary value problems for such models can be solved explicitly using generalized Fourier series (Appendix \ref{app:solDN}).

In a more general case when the helical fiber families can have a radial projection, a modified-fiber wave model \eqref{G:4N} was obtained, containing additional nonlinear terms (Section \ref{sec:4:modif:fibers}). Another nonlinear extension in Section \ref{sec:visc} was obtained by taking into account viscoelastic effects, resulting in the PDEs of the form \eqref{Geq:visco} involving third-order space-time derivatives acting as diffusive terms.

A forced-boundary Neumann problem for dimensionless forms of the linear cylindrical wave equation \eqref{G:lin} and its two nonlinear extensions \eqref{G:4N} and \eqref{Geq:visco} was solved numerically for several sample cases. Simulations indicate singularity formation in modified-fiber wave model \eqref{G:4N}, and regularizing diffusion-type effect of the viscoelastic terms in the PDEs \eqref{Geq:visco}.

It is of interest to further analyze mathematical properties and solution behaviour of the new nonlinear variable-coefficient wave equations \eqref{G:for:genW:f}, \eqref{G:4N}, and \eqref{Geq:visco}, in particular, their conservation laws, solution existence and stability conditions, possibility of construction of closed-form exact solutions, optimal numerical methods, etc. In particular, while the Cartesian constant-coefficient linear wave equation $u_{tt}=c^2 u_{xx}$ admits general, exact d'Alembert (bidirectional traveling wave) solutions, cylindrical wave equations considered in the current work do not admit such solutions even in the linear case. Traveling wave-type exact solutions are known to arise in many other contexts for a variety of more complex models, including solitons and solitary waves for integrable and non-integrable shallow water equations, and exact solutions for two-layer linear and nonlinear media with slow or instant transition \cite{bluman1988exact, il2009d}. Such exact and approximate closed-form solutions may be systematically sought for nonlinear wave models derived in the current contribution using, for example, the Lie symmetry framework and its extensions \cite{BCABook}.

Another important question that requires further study is the applicability of
the models \eqref{G:for:genW:f}, \eqref{G:4N}, \eqref{Geq:visco} to the description of actual fiber-reinforced materials, in particular, biological membranes. Indeed, while the above PDEs have rather simple and short forms, taking into account more realistic constitutive relations \eqref{new_W_gen}, pre-stressed configurations, variable density, compressibility, and other factors, would lead to significantly more mathematically complex wave models.




%
%
%

\subsection*{Acknowledgements}

The authors are thankful to NSERC of Canada for support through the Discovery grant program.

{\footnotesize
\bibliography{CLN_waves_biblio21}
\bibliographystyle{ieeetr}
}

\begin{appendix}

\section{Nonlinear wave equations for a general constitutive relation}\label{app:genW}

In this section, we generalize the results of Section \ref{Sec:HyperelasticLinearModel} by deriving the wave equations governing the shear deformations \eqref{Eq:HyperelasticDeformationClass} of an incompressible anisotropic fiber-reinforced hyperelastic solid involving two helical fiber families \eqref{Eq:HelicalFiberFamily12}, and described by a general constitutive relation
$W^h=W^h_{\rm iso}+W^h_{\rm aniso}$ \eqref{new_W_gen}
involving arbitrary sufficiently smooth functions
$W^h_{\rm iso} =U(I_1,I_2)$ \eqref{eq:constit:iso} and $W^h_{\rm aniso}=V(I_4, I_5, I_6, I_7, I_8)$ \eqref{eq:Eaniso:2fib:gen}.

For the deformation class \eqref{Eq:HyperelasticDeformationClass}, the incompressibility condition $J= \det {\tens{F}} = 1$ is identically satisfied (see \eqref{eq:our:defclass:F}), and the specific forms of invariants $I_j$, $j=1,..,8$, are given by
\beq\label{eq:genW:Ij}
\barr
\mathcal{C}: & I_1=I_2=3+G_R^2,\quad I_3=J^2=1,\quad I_4=I_6=1,\\[2ex] ~& I_5=I_7=1+G_R^2 \sin^2\beta,\quad I_8=\cos^2 2\beta.
\earr
\eeq
The equations of motion for the displacement $G(t,R)$ are derived componentwise from \eqref{eq:PK1:from:Wh}, \eqref{eq:motion:basic:lagr:mom} (see also Figure \ref{fig:test2}). As in Section \ref{Sec:HyperelasticLinearModel},
the $R-$ and $\Phi-$projections of the equation of motion \eqref{eq:motion:basic:lagr:mom} determine the pressure; they are compatible when
\beq\label{eq:genW:Pcond}
\dfrac{\partial V}{\partial I_5}\big|_\mathcal{C}=\dfrac{\partial V}{\partial I_7}\big|_\mathcal{C}.
\eeq
i.e., the indicated partial derivatives match when the substitution $\mathcal{C}$ has been made. The pressure equation then becomes
\beq\label{eq:genW:Peq}
p_R = 4\rho G_R G_{RR}(U_{1,1}+3U_{1,2}+2U_{2,2}) - \dfrac{2\rho}{R} {G_R^2} U_2 - \dfrac{2\rho}{R}\cos^2\beta( V_4+4V_5+V_6+ V_8 \cos 2\beta)
\eeq
(cf. \eqref{Eq:HHEP}). Here and below we denote
\[
V_4=\dfrac{\partial V}{\partial I_4}\big|_\mathcal{C}, \quad U_{1,2}=\dfrac{\partial U}{\partial I_1\partial I_2}\big|_\mathcal{C},
\]
etc., i.e., partial derivatives of general constitutive functions by their respective arguments, computed on relations \eqref{eq:genW:Ij}. Importantly, the substitution $\mathcal{C}$ is made after computing the indicated partial derivatives.

The $Z-$projection of \eqref{eq:motion:basic:lagr:mom} leads to the equation of motion
\beq\label{Eq:Geq:genW}
G_{tt} = \dfrac{1}{R}\dfrac{\partial }{\partial R}\left(2\, R\, (U_1+U_2+2V_5\sin^2\beta)\, G_R \right),
\eeq
governing the nonlinear dynamics of the anti-plane shear displacement $G(t,R)$ (cf. \eqref{G:lin}). The PDE \eqref{Eq:Geq:genW} is briefly written as \eqref{G:lin} in the Introduction.

\section{Exact solutions of the linear wave problems with two families of helical fibers}\label{app:solDN}

\subsection{The Dirichlet problem}

The non-homogeneous Dirichlet initial-boundary value problem \eqref{G:lin}, \eqref{eq:linw:Diri} describes shear radial waves in an elastic cylinder reinforced with two families of fibers (Section \ref{Sec:HyperelasticLinearModel}).

The solution proceeds by the separation of variables in the PDE \eqref{G:lin}: $G(t,R)=T(t)Q(R)$. Considering a homogeneous version of the problem first, one gets
\[
\dfrac{T''}{\alpha T} = \dfrac{R^2Q''+RQ'}{R^2 Q} = -\lambda,\quad \lambda=\const\in \mathbb{R}.
\]
Consequently, the spatial part of the solution $Q=Q(R)$ satisfies the BVP
\begin{equation}\label{Eq:Dir:QBVP}
\begin{array}{lll}
-(RQ')' = \lambda R Q, \quad R_1\leq R\leq R_2;\\[1ex]
Q(R_1) = Q(R_2) = 0.
\earr
\end{equation}
Here the ODE is related to the zeroth-degree Bessel equation, and the problem \eqref{Eq:Dir:QBVP} is a regular Sturm-Liouville eigenvalue problem for the eigenpairs $\{\lambda_n, Q_n(R)\}_{n=1}^\infty$. It follows that there is a countable infinite set of eigenpairs, with eigenfunctions given by
\begin{equation}\label{Eq:Dir:QBVP:sol}
Q_n(R)= \dfrac{-Y_0(\sqrt{\lambda_n} R_2)}{J_0(\sqrt{\lambda_n} R_2)} J_0(\sqrt{\lambda_n} R) + Y_0 (\sqrt{\lambda_n} R)\, , \quad n=1,2,3,\ldots,
\end{equation}
in terms of Bessel functions of order zero. In the appropriate function space, the functions \eqref{Eq:Dir:QBVP:sol} form an orthogonal with respect to the inner product
\beq\label{appx:innerp}
(f,g)=\int_{R_1}^{R_2} f(R) g(R) R \,dR.
\eeq
The eigenvalues $\lambda=\lambda_n$ are members of the increasing positive sequence of roots of the equation
\begin{equation}\label{Eq:Dir:lameq}
J_0(\sqrt{\lambda} R_1) Y_0(\sqrt{\lambda} R_2) - Y_0(\sqrt{\lambda} R_1) J_0(\sqrt{\lambda} R_2)=0,
\end{equation}
in particular, $\lambda_n=\mathcal{O}(n^2)\to \infty$ as $n\to \infty$.

\medskip
Next, the homogeneous Dirichlet IBVP \eqref{G:lin}, \eqref{eq:linw:Diri} is converted to one with zero boundary conditions, but a non-homogeneous PDE. This is achieved with a change of variables
\[
G(t,R)= U(t,R) + V(t,R), \qquad V(t,R)=\dfrac{R-R_2}{R_1-R_2}g(t).
\]
The new unknown $U(t,R)$ then satisfies the problem
\beq\label{eq:linw:Diri:U}
\barr
U_{tt} = \alpha \left(\dfrac{1}{R}U_R + U_{RR}\right) + F(t,R),\quad R_1\leq R\leq R_2,\quad t>0;\\[2ex]
U(R,0) = -V(R,0),\quad U_t(R,0) = -V_t(R,0),\quad R_1\leq R\leq R_2;\\[2ex]
U(R_1,t) =U(R_2,t) = 0, \quad t>0.
\earr
\eeq
Here $F(t,R) :=  ({\alpha}/{R})V_R - V_{tt}$ is a known function in terms of $R_1, R_2, R$, and  $g(t)$.
The problem \eqref{eq:linw:Diri:U} is solved by expanding the forcing term $F(t,R)$ and the unknown $U(t,R)$ in the eigenfunction basis \eqref{Eq:Dir:QBVP:sol}, finding ODEs satisfied by time-dependent coefficients, and solving them, which is a standard procedure (see, e.g., Ref.~\cite{Zauderer}).

We now give an explicit expression for the series solution $G(t,R)$ of the Dirichlet problem \eqref{G:lin}, \eqref{eq:linw:Diri} in the specific case of a periodically driven inner cylinder wall,
\[
g(t)=G_0\sin{\Omega t}, \qquad \Omega=\const>0.
\]
The unique explicit solution for the shear displacements is given by the Fourier series
\beq\label{eq:AppA:DirSolG:Spec}
\barr
G(t,R) = \left(\dfrac{R-R_2}{R_1-R_2}\right) G_0 \sin \Omega t  \\[2ex]
\qquad + \displaystyle \sum_{n=1}^{\infty} \left(\tilde{a}_n \sin(\sqrt{\alpha \lambda_n} t) + \tilde{f}_n \left(\dfrac{\Omega \sin(\sqrt{\alpha \lambda_n} t) - \sqrt{\alpha \lambda_n} \sin\Omega t}{\Omega^2 - \alpha \lambda_n} \right) \right) Q_n(R),
\earr
\eeq
where $Q_n(R)$ is given by \eqref{Eq:Dir:QBVP:sol}, eigenvalues $\lambda_n$ satisfy \eqref{Eq:Dir:lameq}, and the Fourier coefficients are
\[
\begin{split}
\tilde{a}_n &= -\dfrac{\Omega G_0}{R_1-R_2} {\displaystyle\int_{R_1}^{R_2} \left(R^2 - R R_2\right) Q_n\, dR}\Big/{\displaystyle\int_{R_1}^{R_2} Q_n^2 R \,dR}, \\[3ex]
\tilde{f}_n &= \dfrac{G_0}{R_1-R_2} {\displaystyle\int_{R_1}^{R_2} \left( \Omega^2 R (R-R_2) + \alpha \right)Q_n \,dR}\Big/{\displaystyle\int_{R_1}^{R_2} Q_n^2 R \,dR}. \\
\end{split}
\]
We note that as expected, the resonance occurs when the forcing frequency coincides with one of the eigenfrequencies: $\Omega^2 = \alpha \lambda_n$.


\subsection{The Neumann Boundary Value Problem}

In a similar fashion, one can construct an explicit series solution to a non-homogeneous Neumann initial-boundary value problem \eqref{G:lin}, \eqref{eq:linw:Neu}. The separation of variables in the PDE \eqref{G:lin}, using $G(t,R)=T(t)Q(R)$, proceeds in the same fashion as for the Dirichlet problem. In the Neumann case, the spatial part of the solution $Q=Q(R)$ satisfies the Sturm-Liouville BVP
\begin{equation}\label{Eq:Neu:QBVP}
\begin{array}{lll}
-(RQ')' = - \lambda R^2 Q, \quad R_1\leq R\leq R_2;\\[1ex]
Q'(R_1) = Q'(R_2) = 0,
\earr
\end{equation}
which yields the eigenpairs $\{\mu_n, \tilde{Q}_n(R)\}_{n=1}^\infty$. The eigenfunctions are given by
\begin{equation}\label{Eq:Neu:QBVP:sol}
\tilde{Q}_n(R)= - \dfrac{Y_1(\sqrt{{\mu_n}} R_2) J_1 (\sqrt{\mu_n} R)}{J_1 (\sqrt{\mu_n} R_2)} + Y_1 (\sqrt{\mu_n} R)
\end{equation}
in terms of Bessel functions of order one, and are orthogonal with respect to the same inner product \eqref{appx:innerp}. The eigenvalues $\mu_n$ are also different from their Dirichlet homologues; they are the members of the increasing positive sequence of roots of the equation
\[
J_1(\sqrt{\mu} R_1) Y_1(\sqrt{\mu} R_2) - Y_1(\sqrt{\mu} R_1) J_1(\sqrt{\mu} R_2)=0.
\]
The rest of the solution procedure carries over from the Dirichlet case above, using expansions in the basis \eqref{Eq:Neu:QBVP:sol}. We again give an explicit expression for the series solution $G(t,R)$ of the Neumann problem \eqref{G:lin}, \eqref{eq:linw:Neu} in the specific case of a periodically forced inner cylindrical wall:
\[
g(t)=G_0\sin{(\Omega t)}, \qquad \Omega=\const>0.
\]
The solution takes the form
\beq\label{eq:AppA:NeumSolG:Spec}
\barr
G(t,R) = \dfrac{R^2-R_2 R}{2(R_1-R_2)} \,G_0 \sin \Omega t \\[2ex]
\qquad + \displaystyle\sum_{n=1}^{\infty} \left(\tilde{a}_n \sin(\sqrt{\alpha \mu_n} t) + \tilde{f}_n \left(\frac{\Omega \sin(\sqrt{\alpha \mu_n} t) - \sqrt{\alpha \mu_n} \sin\Omega t}{\Omega^2 - \alpha \mu_n} \right) \right) \tilde{Q}_n(R),
\earr
\eeq
where the Fourier coefficients are given by
\[
\begin{split}
\tilde{a}_n &= \left(\dfrac{\Omega G_0}{2(R_1-R_2)}\right) {\int_{R_1}^{R_2} \left(R^3 - 2 R^2 R_2\right) \tilde{Q}_n \,dR}\Big/{\int_{R_1}^{R_2} \tilde{Q}_n^2 R \,dR}, \\[3ex]
\tilde{f}_n &= \dfrac{G_0}{2(R_1-R_2)} {\int_{R_1}^{R_2} \left( \Omega^2 R^3 - 2 \Omega^2 R_2 R^2 + 4 \alpha R - 2 \alpha R_2 \right)\tilde{Q}_n \,dR}\Big/{\int_{R_1}^{R_2} \tilde{Q}_n^2 R\, dR}.
\end{split}
\]

\end{appendix}

\end{document}